\documentclass[pdftex,twocolumn,epjc3]{svjour3}  

\RequirePackage[T1]{fontenc}

\RequirePackage{graphicx}
\RequirePackage{mathptmx}      
\RequirePackage{flushend}
\RequirePackage[numbers,sort&compress]{natbib}
\RequirePackage[colorlinks,citecolor=blue,urlcolor=blue,linkcolor=blue]{hyperref}

\usepackage{amsmath}

\journalname{Eur. Phys. J. A}
\begin{document}


\title{Quasi-exotic open-flavor mesons\thanksref{t1}}

\author{T. Hilger\thanksref{e1,addr1}
        \and
        A. Krassnigg\thanksref{e3,addr1}
        }

\thankstext{t1}{This work was supported by the Austrian Science Fund (FWF) under project no.\ P25121-N27.}
\thankstext{e1}{e-mail: thomas.hilger@uni-graz.at}
\thankstext{e3}{e-mail: andreas.krassnigg@uni-graz.at}


\institute{Institute of Physics, University of Graz, NAWI Graz, A-8010 Graz, Austria \label{addr1}
}

\date{Received: \today / Accepted: date}

\maketitle

\abstract{Meson states with exotic quantum numbers arise naturally in a covariant bound-state
framework in QCD. We investigate the consequences of shifting quark masses such that
the states are no longer restricted to certain $\mathcal{C}$-parities, but only by $J^\mathcal{P}$. Then,
\emph{a priori}, one can no longer distinguish exotic or conventional states. In order
to identify signatures of the different states to look for experimentally, we provide the 
behavior of masses, leptonic decay constants, and orbital-angular-momentum decomposition
of such mesons, as well as the constellations in which they could be found. Most
prominently, we consider the case of charged quasi-exotic excitations of the pion.
}


\maketitle

\section{Introduction\label{sec:intro}}


The appeal of discovering the patterns of mesons with exotic quantum numbers \cite{Meyer:2010ku}
comes from the same source as the term \emph{exotic} itself, namely the quark-model
picture \cite{Eichten:1979ms,Godfrey:1985xj,Eichten:1994gt}. 
Mesons are classified via their total angular momentum $J$,
parity $\mathcal{P}$, and, if the state can be seen as its own antiparticle, by charge-conjugation
parity $\mathcal{C}$. In the quark-model setup, one combines one quark and one antiquark with
total spin $s$ and orbital angular momentum $l$ to get a meson with $J^{\mathcal{P}\mathcal{C}}$, where
$\mathcal{P}=(-1)^{l+1}$, $\mathcal{C}=(-1)^{l+s}$, and $|l-s|\le J \le |l+s|$. 
Any combination of $J^{\mathcal{P}\mathcal{C}}$ that violates these constraints is termed \emph{exotic}.

Indeed, the patterns
of meson states predicted by the quark model matched those found in experiment. 
Exotic sta\-tes do not seem to appear in the experimental
meson spectrum with the exception of the isovector $1^{-+}$ case 
\cite{Alde:1988bv,Aoyagi:1993kn,Thompson:1997bs,Abele:1998gn,Abele:1999tf,Dzierba:2003fw,Salvini:2004gz,Adams:2006sa,
Alekseev:2009aa,Lu:2004yn,Kuhn:2004en,Ivanov:2001rv,Adams:1998ff,Dzierba:2005jg,
Lu:2004yn,Kuhn:2004en
}. 
Since even those states are under debate, both systematic
experimental evidence for states with exotic quantum numbers as well as a sound theoretical
concept and understanding of their peculiarities are among the top priorities of
modern hadron physics \cite{Briceno:2015rlt}.

Still, it is puzzling that there should be such a profound
conceptual difference between exotic and conventional me\-sons from the point of view
of a covariant bound-state amplitude, since there the above restrictions do not apply.
In particular, a Poincar\'{e} covariant formulation of the two-body bound-state
problem additionally has a relative time freedom of the constituents and thus
lifts the nonrelativistic $J^{\mathcal{P}\mathcal{C}}$ limitations \cite{Burden:2002ps}.
For a quark bilinear Bethe-Salpeter amplitude (BSA), one has three four-vectors
and their scalar products as building blocks: the total momentum $P$, the $\bar{q}q$ relative 
momentum $k$, and the four vector of Dirac matrices $\gamma$ as a representation
of the direct spinor product \cite{Hu:1966ab,Hu:1966bs,Smith:1969az}.

For example, a scalar BSA has four possible independent covariants:
\begin{equation}\label{eq:tbase}
\texttt{t}_j\in \{\mathbf{1},\;\gamma\cdot P,\;\gamma\cdot k,\;-\frac{1}{2}\,[\gamma\cdot k,\gamma\cdot P]\}_{j=1,2,3,4}\;,
\end{equation}
where $\mathbf{1}$ is the unit matrix 
in Dirac space. The BSA for $J^\mathcal{P}=0^+$ is obtained as  
\begin{equation}\label{eq:bsa}
\Gamma(k;P;\gamma):=\sum^4_{i=1}\texttt{t}_i(k;P;\gamma)\; F_i(k^2,k\cdot P,P^2)\;,
\end{equation}
the generalization to other $J^\mathcal{P}$ is straightforward \cite{Krassnigg:2010mh}. 
The invariant amplitudes $F_i$ are parameterized in terms of the
Lorentz-invariants $P^2$, $k^2$, and $k\cdot P$. 
There is no \emph{a priori} restriction of $\mathcal{C}$ at this point.
While each covariant $\texttt{t}_i$ has a definite $\mathcal{C}$-parity, the dependence
of the $F_i$ on $k\cdot P$, which has $\mathcal{C}=-1$ in contrast to $P^2$ and $k^2$, allows for 
both $\mathcal{C}=\pm 1$ of the BSA. The construction of Eq.~(\ref{eq:bsa}) is also valid for
the case of an open-flavor meson \cite{Maris:1997tm,UweHilger:2012uua,Fischer:2014xha,Rojas:2014aka,Leitao:2016bqq,Hilger:2016drj,Hilger:2017jti}, 
where the $F_i$ do not possess a definite symmetry
regarding $k\cdot P$. A covariant approach thus allows for any set of
$J^{\mathcal{P}\mathcal{C}}$ without the explicit appearance of gluonic or other degrees of freedom
additional to quark bilinears \cite{Burden:1996nh,Watson:2004kd,Krassnigg:2009zh,Fischer:2014cfa,Hilger:2015ora}.

The most common starting point to address an exotic state is the notion
of \emph{hybrid} mesons, which \emph{explicitly} contain some gluonic excitation. Hybrid-meson
supermultiplets contain states with both exotic and conventional quantum numbers and have different content 
and properties, depending on the method of investigation. In a recent article \cite{Dudek:2011bn}, lattice 
results \cite{Dudek:2010wm,Dudek:2013yja} were partly contrasted to 
various different model setups \cite{Horn:1977rq,Isgur:1984bm,Jaffe:1985qp,Barnes:1982tx,Chanowitz:1982qj,General:2006ed,Guo:2008yz}. 
The results for mass ranges of the hybrid 
supermultiplets described there vary as does the candidate for the lightest meson with 
exotic quantum numbers. The majority of investigations finds them close to or above $2$ GeV, 
and the lightest is, e.\,g., the $J^{\mathcal{P}\mathcal{C}}=1^{-+}$ \cite{Chen:2013zia}, or the $0^{--}$ \cite{Cotanch:2010bq}.

In the Dyson-Schwinger-Bethe-Salpeter-equation
(DSBSE) approach, exotic mesons have been studied in detail some time ago in the context of
the $1^{-+}$ channel \cite{Burden:2002ps}, which was also part of our recent line of 
investigations \cite{Blank:2011ha,Popovici:2014pha,Hilger:2014nma,Hilger:2015hka}, and for $J=0,1$ in \cite{Qin:2011xq}. 
Still, herein we present the first systematic set of predictions and conclusions for such states
in our approach, which are anchored to some of QCD's model-independent properties, in particular the behavior of the iso\-vector pseudoscalar meson ground-state mass and leptonic decay constant as well as the leptonic decay constants of all its excitations in the chiral limit. 
The corresponding excited-state masses, on the other hand, are not anchored to QCD properties, since they are not constrained by the axial-vector Ward-Takahashi identity. This provides freedom and predictive power for the model calculation in the sense that these masses vary strongly with changes in some model parameters and can be used to test and falsify our results. 

Irrespective of the approach and the construction with respect to gluonic degrees of freedom used to describe me\-sons, one faces a very interesting
situation in the open-flavor case. Since such states are not restricted by $\mathcal{C}$, one needs
criteria, signals, or simply hints as to whether some of these states have exotic characteristics,
i.\,e., do not exist in the quark model, and which ones.
Clearly, one can expect some kind of similarity or correspondence of open-flavor states and certain quarkonia.

\begin{figure}[t]
\centering
  \includegraphics[width=.4\textwidth]{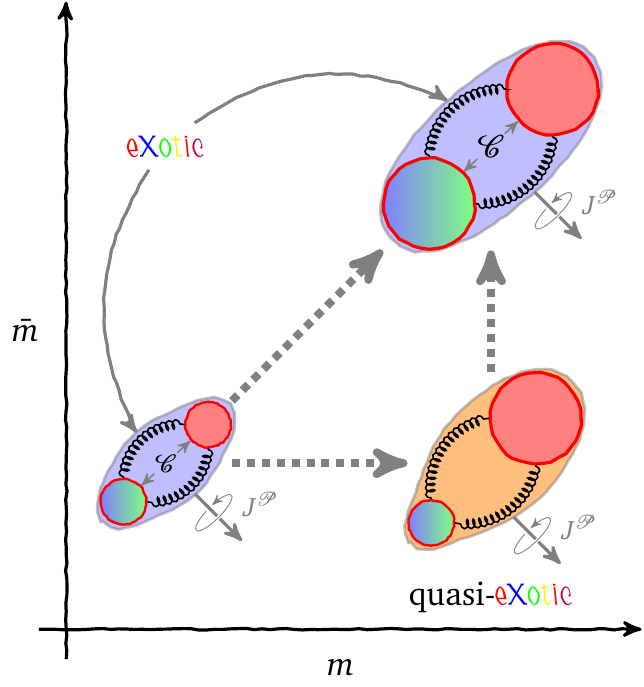}
\caption{\label{fig:quasiexotic} Schematics of the connection of exotic and quasi-exotic mesons.}
\end{figure}

Concretely, as illustrated in Fig.~\ref{fig:quasiexotic}, one can approach the case 
of a $\mathcal{C}$-eigenstate from the perspective of a meson with strangeness $\bar{n}s$,
both towards the $\bar{n}n$ and $\bar{s}s$ limits. This is based on the reasonable notion that, when varying quark masses,
mass trajectories of corresponding meson bound states should be continuous and no states should disappear or appear. 
We use the term \emph{quasi-exotic} for flavored mesons that can be connected to 
exotic $\mathcal{C}$-eigenstates. We argue that such states do not exist in the quark model, 
similar to exotic states. As it turns out, the most prominent and probably most characteristic
instance of a quasi-exotic meson could be an excited charged pion. 

Note that mesons in QCD are strong resonances above the respective decay thresholds. 
While this is a qualitative difference to the bound-state picture used herein as a result of the fact that there are no strong decay mechanisms in the truncated used as described below, we do not expect a model calculation to be inexplicably discontinuous as a function of the current-quark mass, even if it does include hadronic decay channels. Thus, our qualitative argument does not suffer from this omission. 
Nevertheless, a treatment of meson states as resonances, where applicable, is a necessary next step in model calculations such as ours.

\section{Properties of mesons in QCD\label{sec:general}}

The Dyson-Schwinger equations (DSEs) of QCD are the equations of motion in this quantum field theory \cite{Fischer:2006ub},
a general concept easily extendable, e.\,g., to finite temperature and density 
\cite{Roberts:2000aa,Blank:2010bz,Horvatic:2010md,Eichmann:2015kfa,Cui:2016zqp,Gao:2016qkh,Gao:2016hks}. 
Their complete solution amounts to a solution of QCD in terms of the dynamics of quarks and gluons; 
however, this task is highly nontrivial \cite{Bender:2002as,Bhagwat:2004hn,Gomez-Rocha:2014vsa,Gomez-Rocha:2015qga,Gomez-Rocha:2016cji,%
Chang:2009zb,Sanchis-Alepuz:2015qra,Sanchis-Alepuz:2015tha}. While numerical studies such as ours require a truncation
of this infinite tower of coupled nonlinear integral equations, some results can be obtained in a
truncation-independent manner, since they are connected to (broken) symmetries of QCD and are 
realized through Ward-Takahashi identities (WTIs) or Slavnov-Taylor identities \cite{Marciano:1977su}.

Such a truncation-independence of meson properties is basically qualitative of nature, but also quantitative in the sense that a value of zero is precisely reproduced and values close to a corresponding point or limit are also quantitatively reliable, if properly anchored via appropriate choices for the model parameters. 
In our case, we concretely have the exact behavior in the chiral limit of meson properties as described above. 
Thus, for small quark masses, our calculated values for the corresponding anchored mass or leptonic decay constant should also be reliable.
The properties and relations considered regarding this feature are described in the following.
Our results are presented with a clear qualitative focus and a quantitative note, where we mention experimental values.

A prime example is the axial-vector WTI which is connected to QCD's chiral symmetry and the both dynamical
and explicit breaking thereof \cite{Maskawa:1974vs,Aoki:1990eq,Kugo:1992pr,Bando:1993qy,Munczek:1994zz,Biernat:2014xaa}. 
If a truncation, such as the RL truncation used here, satisfies this WTI,
all aspects of chiral symmetry and its dynamical breaking are realized in the model by construction. In 
particular, the pion ground state becomes massless in the chiral limit, where it is identified as the 
Goldstone boson connected to dynamical chiral symmetry breaking (DCSB). Further well-known relations appear
as consequences of WTIs, such as the Gell-Mann-Oakes-Renner relation. A generalized form of the latter \cite{Maris:1997hd}
can be written together with a similar relation for scalar mesons \cite{Qin:2011xq} as
\begin{equation}\label{eq:ggmor}
f_{0^\mathcal{P}}\; M_{0^\mathcal{P}}^2=(m_q-\mathcal{P}\; m_{\bar{q}})\; r_{0^\mathcal{P}}\;,
\end{equation}
which provides both model independent insight as well as a means to check numerical studies. 
$f_{0^\mathcal{P}}$ and $M_{0^\mathcal{P}}$ are the 
meson's leptonic decay constant and mass, respectively,
$m_{(\bar{q})q}$ are the current-(anti)quark masses, and $r_{0^\mathcal{P}}$ is a projection of, e.\,g., the pseudoscalar BSA
on a pseudoscalar current instead of the axialvector that defines $f_{0^-}$ \cite{Maris:1997tm}.

For pseudoscalar mesons it is known \cite{Dominguez:1976ut,Dominguez:1977nt,Volkov:1999yi} and established also 
in the DSBSE approach \cite{Holl:2004fr} via Eq.~(\ref{eq:ggmor}) that leptonic decay constants of all radially 
excited pseudoscalar mesons are zero in the chiral limit, if one has DCSB. For realistic 
light-quark masses one arrives at values of the order of $f_\pi\approx 1-10$ MeV
for the first radial excitation of the pion both in the DSBSE and other approaches 
\cite{Holl:2003dq,Holl:2004fr,McNeile:2006qy,Mastropas:2014fsa,Ballon-Bayona:2014oma,Krein:2016gua}. 
While all radially excited pseudoscalar mesons have a small leptonic decay constant compared to the pion
ground state's, exotic $0^{--}$ excitations have a vanishing decay constant as
a direct result of the projection onto the axialvector current.

For scalar mesons, it is evident from Eq.~(\ref{eq:ggmor}) that the projection $f_{0^+}$ of a scalar BSA on a vector
current vanishes for all states with $m_q=m_{\bar{q}}$ irrespective of their level of excitation, including exotic $0^{+-}$ quantum
numbers \cite{Bhagwat:2006py,Qin:2011xq}. For vector mesons their leptonic decay constant $f_{1^-}$ 
\cite{Maris:1999nt} is generally nonzero for conventional and zero for exotic states. 
Our numerical investigation is based on these boundary conditions, which are satisfied
to high accuracy.

\section{Essentials of model setup\label{sec:model}}
RL truncation combines the rainbow truncation of the quark DSE with the ladder truncation
of the $\bar{q}q$ BSE via an effective model quark-gluon interaction \cite{Maris:1997tm,Maris:1999nt}. 
First, the quark DSE is solved numerically and prepared as an input to the meson BSE 
\cite{Krassnigg:2008gd,Krassnigg:2010mh,Dorkin:2010ut,UweHilger:2012uua,Dorkin:2013rsa}, which
is then solved numerically as well via well-established techniques \cite{Krassnigg:2003wy,Blank:2010bp,Hilger:2015zva}. 
The quark dressing functions follow typical patterns of DCSB in QCD
\cite{Holl:2004un,Krassnigg:2016hml}.

\begin{figure*}[t]
  \includegraphics[width=.32\textwidth]{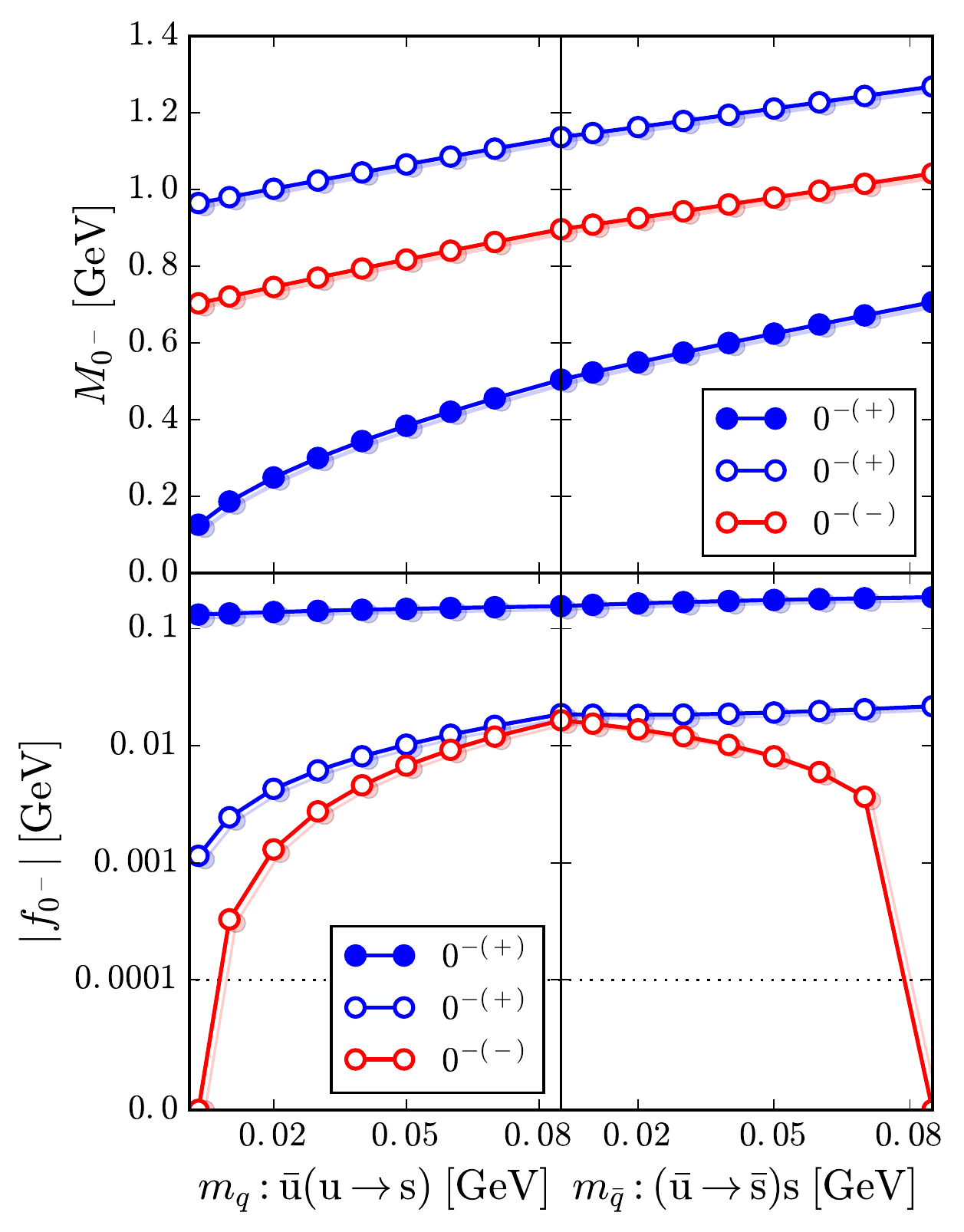}
  \includegraphics[width=.32\textwidth]{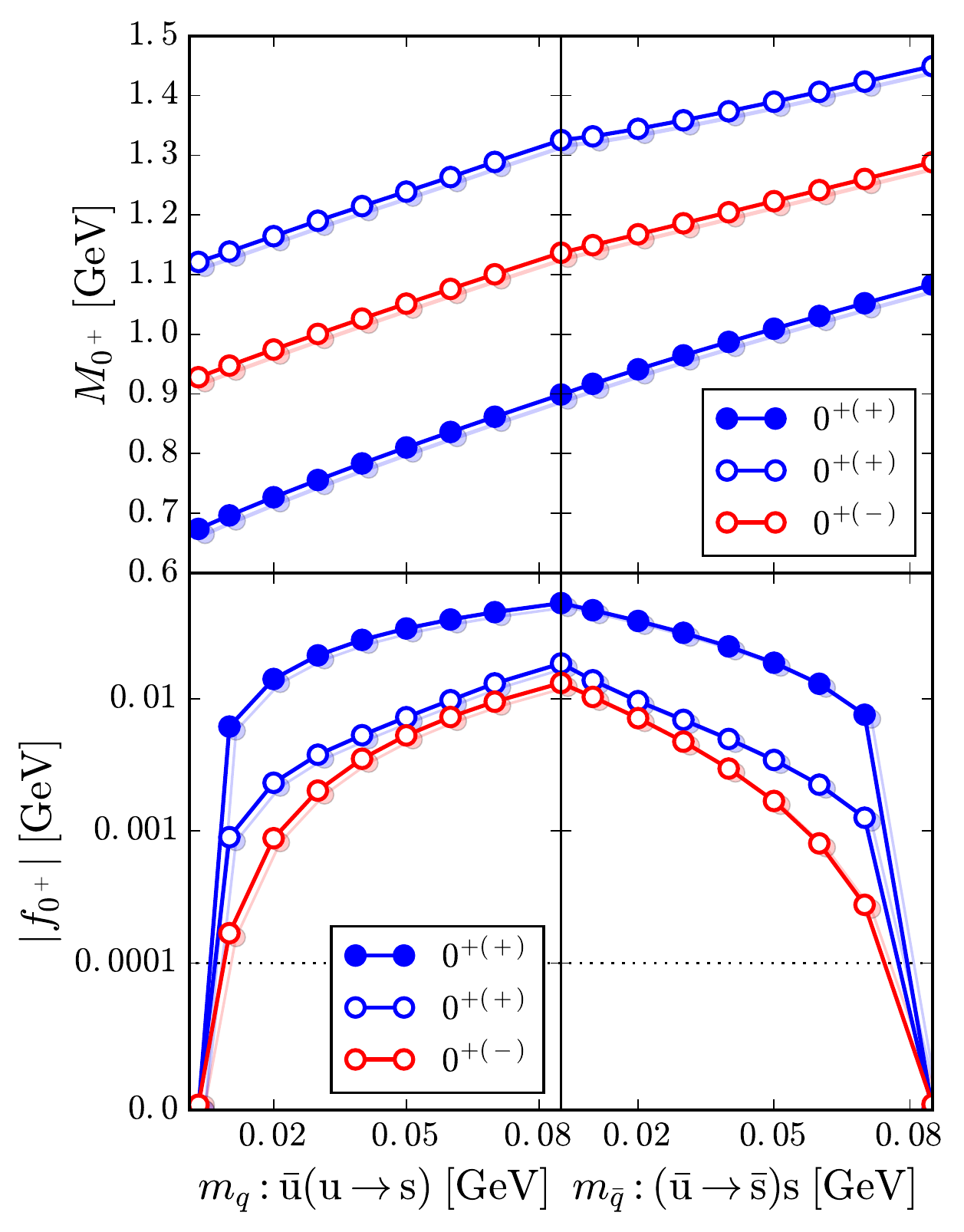}
  \includegraphics[width=.32\textwidth]{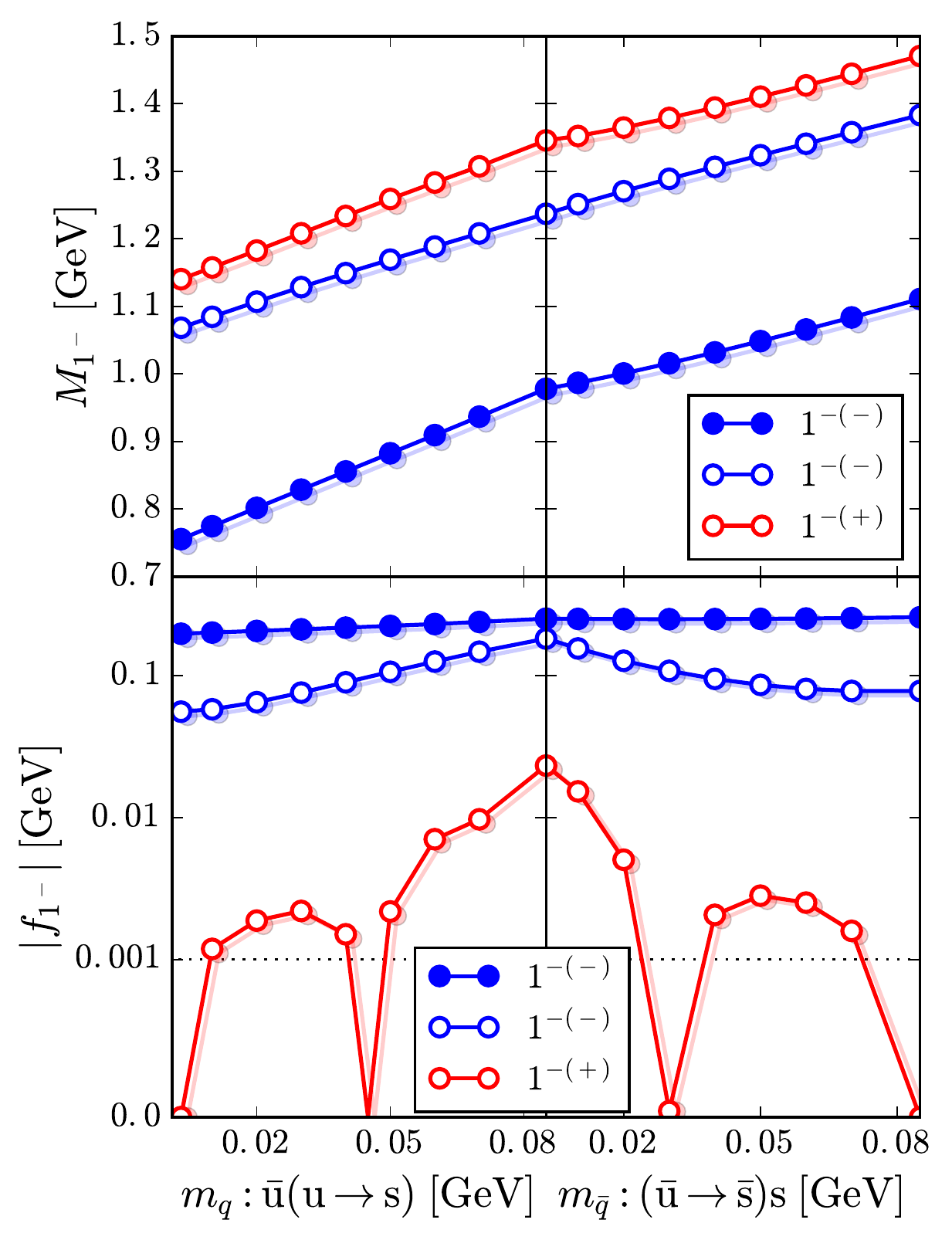}
\caption{\label{fig:mf}
Masses (\emph{top}) and decay constants (\emph{bottom}) as functions of quark masses; left half of each figure: quark mass increases
from light to strange; right half of each figure: antiquark mass increases from light to strange. 
\emph{From left to right:} pseudoscalar, scalar, and vector. Below the dotted line,
the vertical axis is linear instead of logarithmic.}
\end{figure*}

Instead of using a Chebyshev expansion for the $F_i$ (for an illustration see 
\cite{Krassnigg:2003dr,Qin:2011xq,Hilger:2015zva}) and having to keep many Chebyshev moments, we retain the full angular dependence in our calculation as it was pioneered and detailed in \cite{Maris:1997tm,Maris:1999nt} together with the usual convergence checks for numerically discretized integration
in order to preserve independence from the $\bar{q}q$ momentum partitioning 
\cite{Alkofer:2002bp} demanded by Poincar\'{e} covariance. Varying the partitioning in open-flavor DSBSE 
calculations \cite{UweHilger:2012uua} allows us to sample quark propagators only on their analytic domain \cite{Bhagwat:2002tx}
and to directly solve the homogeneous BSE, obtaining the masses and leptonic decay constants
shown in Fig.~\ref{fig:mf}.
Note that exotic meson masses are underestimated in RL truncation if compared to other approaches,
but are shifted up by corrections beyond RL, as one must expect, see \cite{Eichmann:2016yit} and references therein.

Unless noted otherwise, we use the model of Ref.~\cite{Maris:1999nt}, which has the correct UV limit of perturbative QCD
and an intermediate-momentum enhancement producing DCSB parameterized by an inverse effective range $\omega$ 
and an overall strength $D$. Our values $\omega=0.3$ GeV, $D=1.3$ GeV${^2}$ are chosen close to one of the original sets \cite{Maris:1999nt}
aimed at our particular purposes of this study: a rich data set and the exact boundary conditions described above.

To ensure high-quality numerical results and reduce the chance of error as much as possible, we leave very little room for manual errors by high automatization of the calculational setup. In particular, we employ configuration-con\-sistency checks, various plausibility and consistency checks at the level of the results, qualitative and quantitative data checks, and use automatic data visualization for a manual data check, when necessary.

\section{Results and discussion\label{sec:results}}

We investigate the $m_q$ and subsequent $m_{\bar{q}}$ evolution of the meson masses and leptonic decay constants of the ground 
state as well as two excited states for $J^\mathcal{P}=0^-$, $0^+$, $1^-$, where one excitation is connected to exotic and 
one to conventional $\mathcal{C}$-eigenstates. The quantum numbers $J^{\mathcal{P}(\mathcal{C})}$ of each state together with a continuity requirement for
the meson masses, decay constants, orbital angular momentum properties, and relation (\ref{eq:ggmor}) are utilized to uniquely identify 
and assign conventional and quasi-exotic open-flavor states to their conventional and exotic $\mathcal{C}$-eigenstate counterparts.

The light and strange quark masses are fitted to the experimental values of the 
$\pi$ and $K$ mesons' leptonic decay constants, $f_{\pi^-} = 0.13041(20)$ and $f_{K^-} = 
0.1562(7)$ GeV \cite{Olive:2014rpp}. At our renormalization point of $\mu=19$ GeV \cite{Maris:1997tm}, 
$m_n(\mu)=0.003$ and $m_s(\mu)=0.085$ GeV, which yield the calculated 
$f_{\pi^-} = 0.132$ and $f_{K^-} = 0.156$ GeV, respectively.

We present our results in Fig.~\ref{fig:mf}. The first excitation in our model setup is exotic (red) and the second
conventional (blue). For the $1^-$ case, we show higher excitations (whose order is reversed), since for the first two continuous
mass curves cannot be produced because between the $\bar{n}n$ and $\bar{s}s$ cases the homogeneous BSE does not
produce real eigenvalues for these states at some point. This is a non-numerical problem encountered typically in open-flavor DSBSE model calculations \cite{Rojas:2014aka} and is under current investigation also for states representing baryons \cite{Eichmann:2016hgl} or with equal-mass constituents \cite{Eichmann:2016nsu}. 
However, this behavior does not pertain to our argument, since our focus is the association of exotic and quasi-exotic states.

The leptonic decay constants are presented in the lower row of Fig.~\ref{fig:mf},
where we plot $|f|$ in each case, since $f$ has alternating positive and negative signs 
for a tower of radial excitations \cite{Holl:2004fr,Bhagwat:2007rj,Blank:2010sn}. The accurate realization of the 
exact boundary conditions given above is perfectly visible for the scalar case (middle), where all values for $f$ drop to zero
in both $\bar{q}q$ limits.

For the pseudoscalars, the situation is a bit more subtle: In the presence of DCSB,
the ground-state pion's decay constant is sizeable and, when increasing one quark mass to the strange
quark's value, rises slowly to the corresponding value for the $K$. The first conventional excitation has
$f\approx 1$ MeV, which is two orders of magnitude smaller than for the ground state and only nonzero as a result
of explicit chiral symmetry breaking by the current light-quark mass. 

The first $\bar{n}n$ exotic pseudoscalar
has an $f$ of exactly zero. Interestingly, if other studies of hybrids with exotic quantum numbers
find a small but finite value for $f$, this provides an excellent way to test such hypotheses against ours.

With one quark mass increasing to $m_s$, both excitations' values for $f$ increase to a number of just
under $20$ MeV, which means that in the strange sector, according to our picture, a kaon excitation expected from the 
quark model would be indistinguishable from the quasi-exotic state we show here on the basis of their leptonic decay constants.

\begin{figure*}[t]
  \includegraphics[width=.32\textwidth]{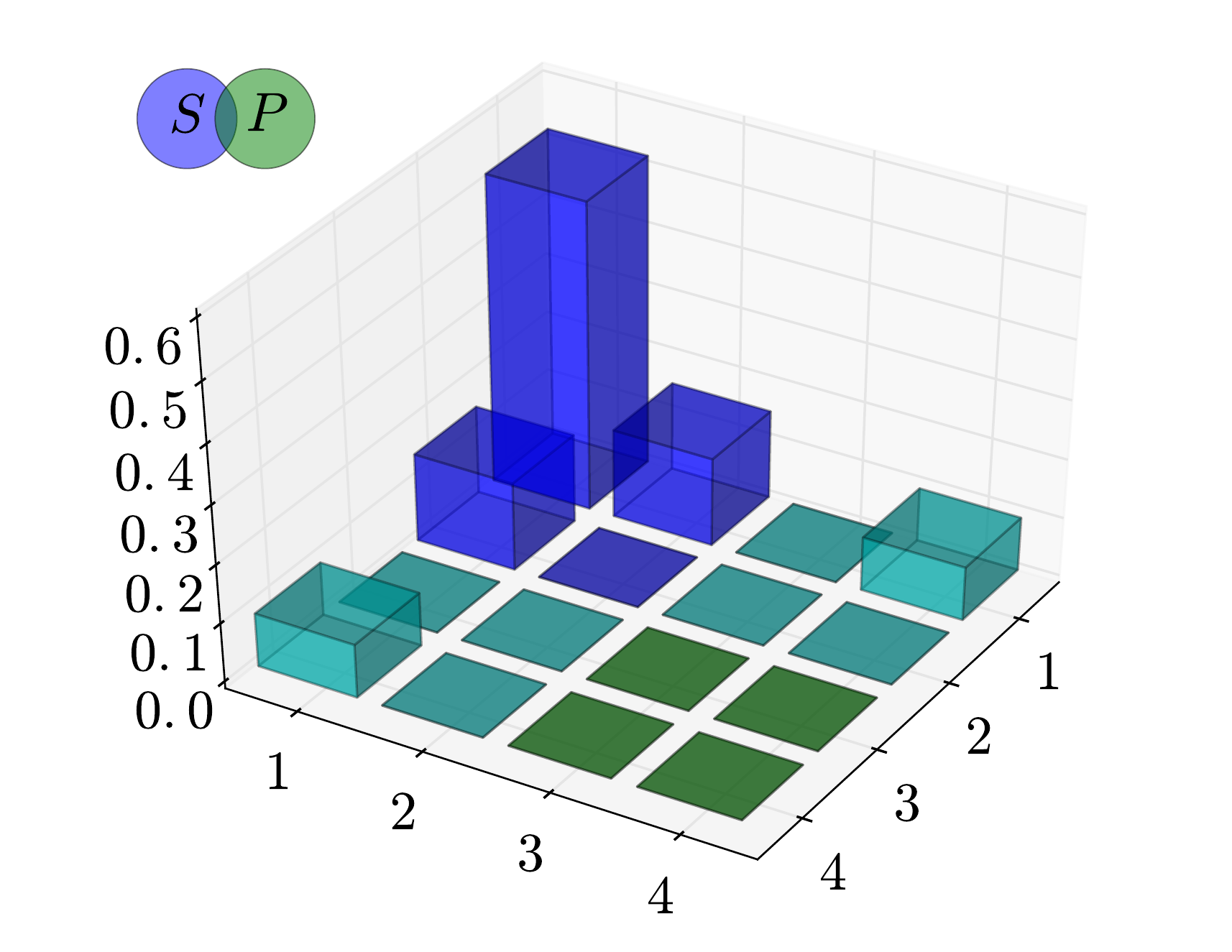}
  \includegraphics[width=.32\textwidth]{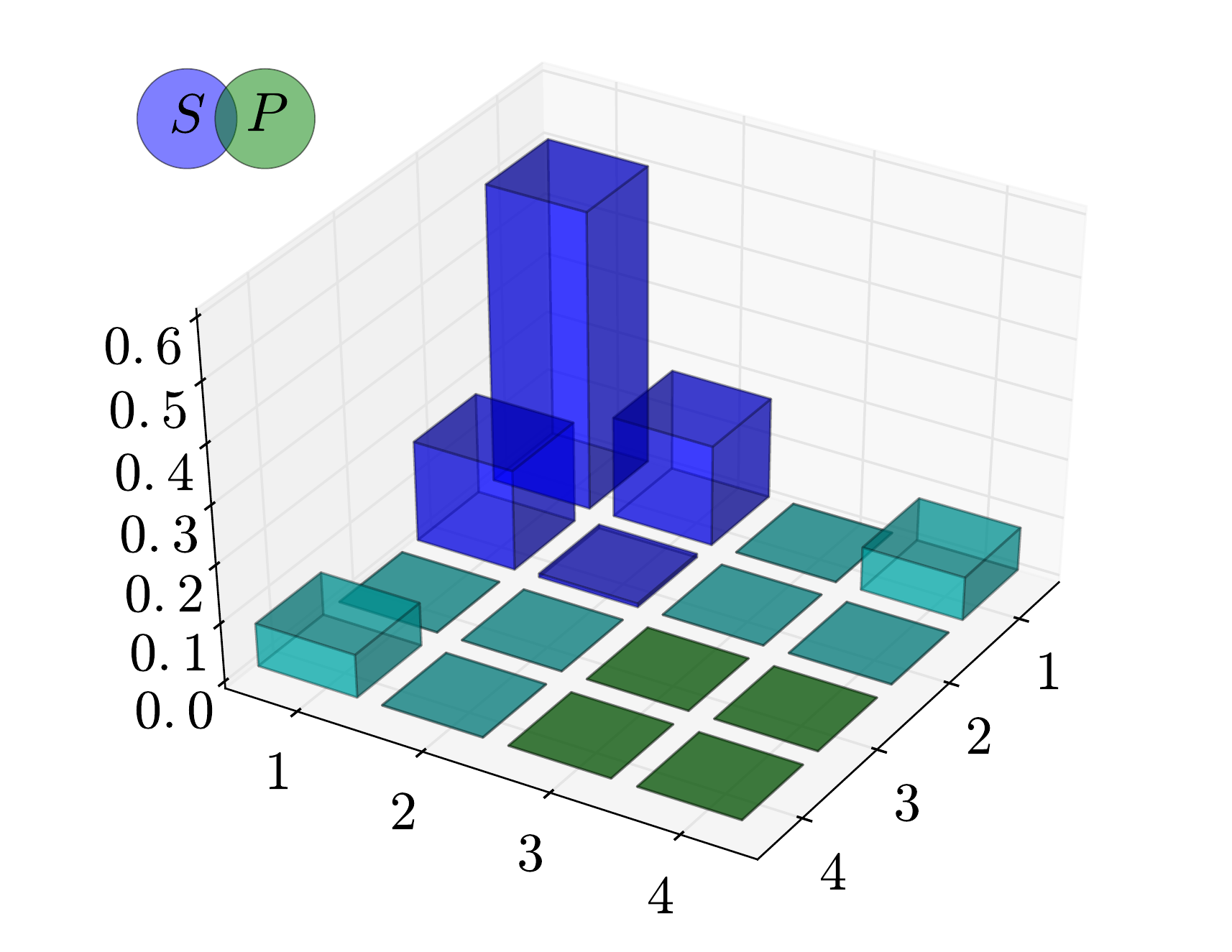}
  \includegraphics[width=.32\textwidth]{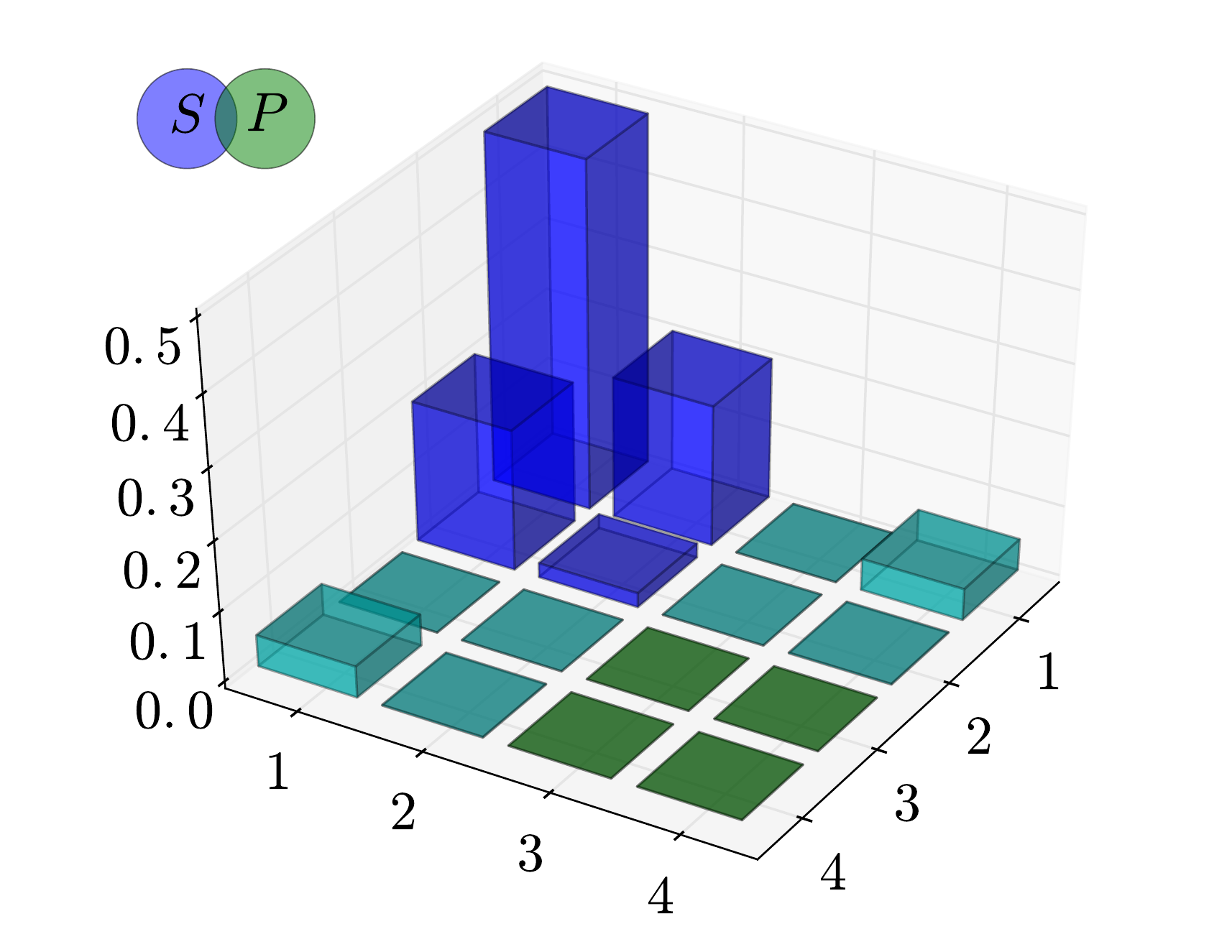}
  \includegraphics[width=.32\textwidth]{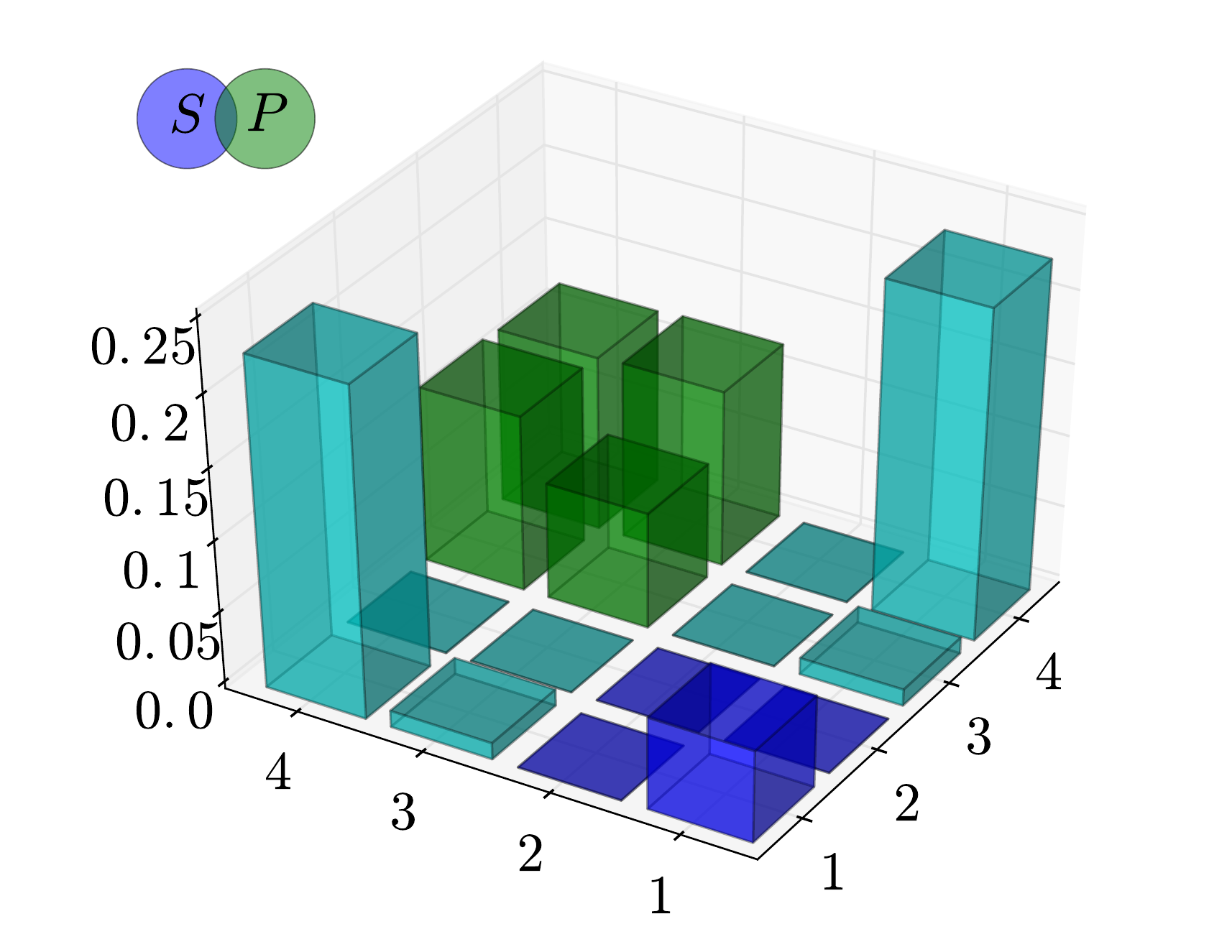}
  \includegraphics[width=.32\textwidth]{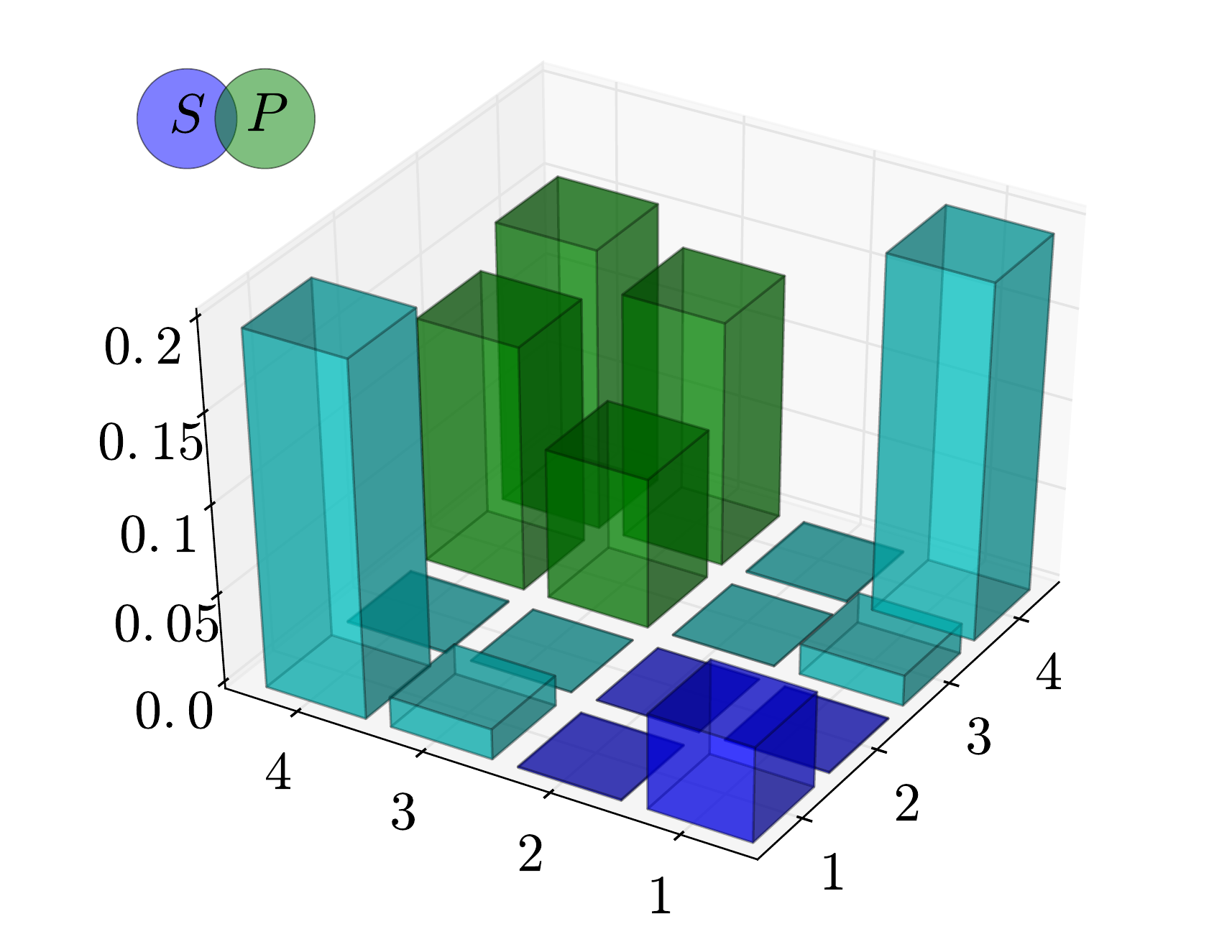}
  \includegraphics[width=.32\textwidth]{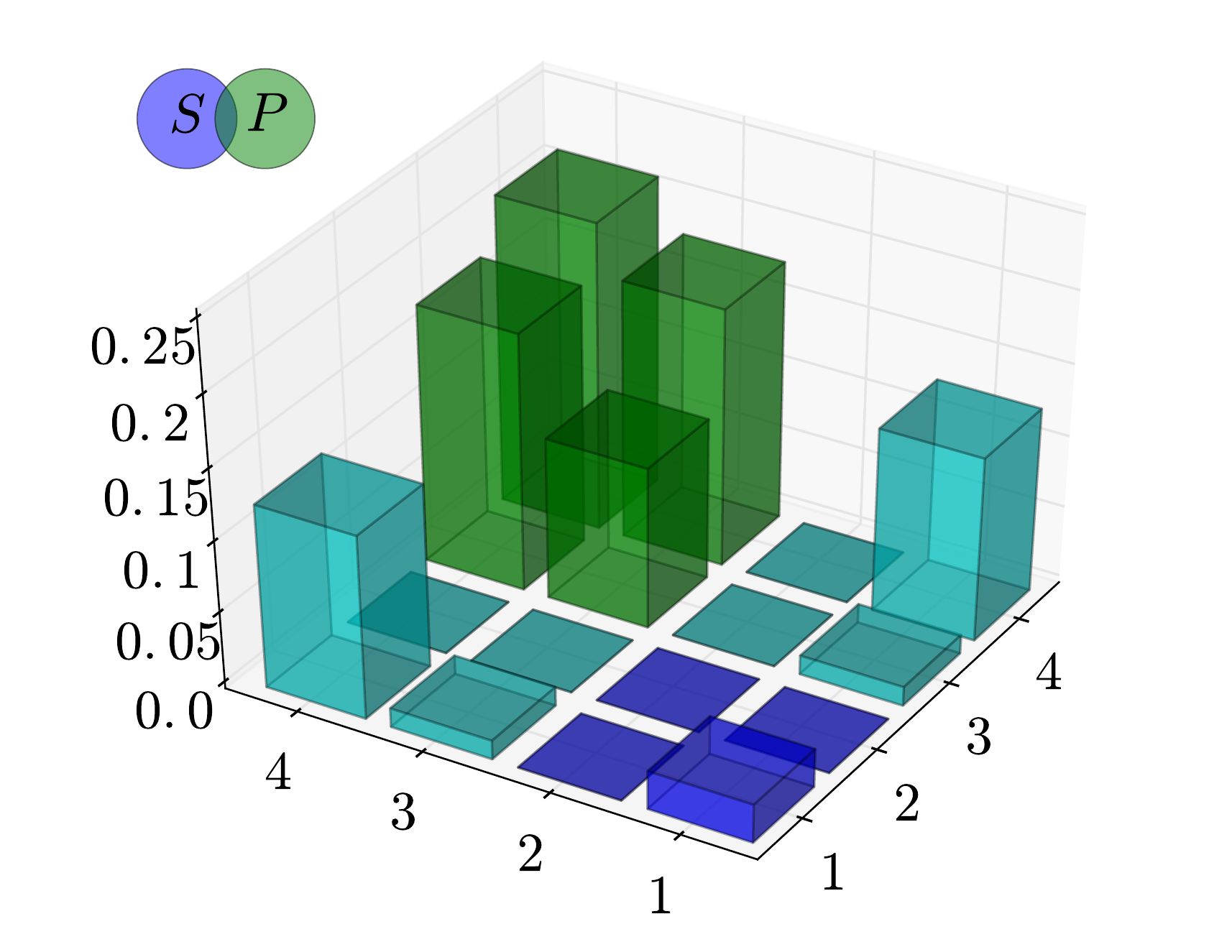}
  \includegraphics[width=.32\textwidth]{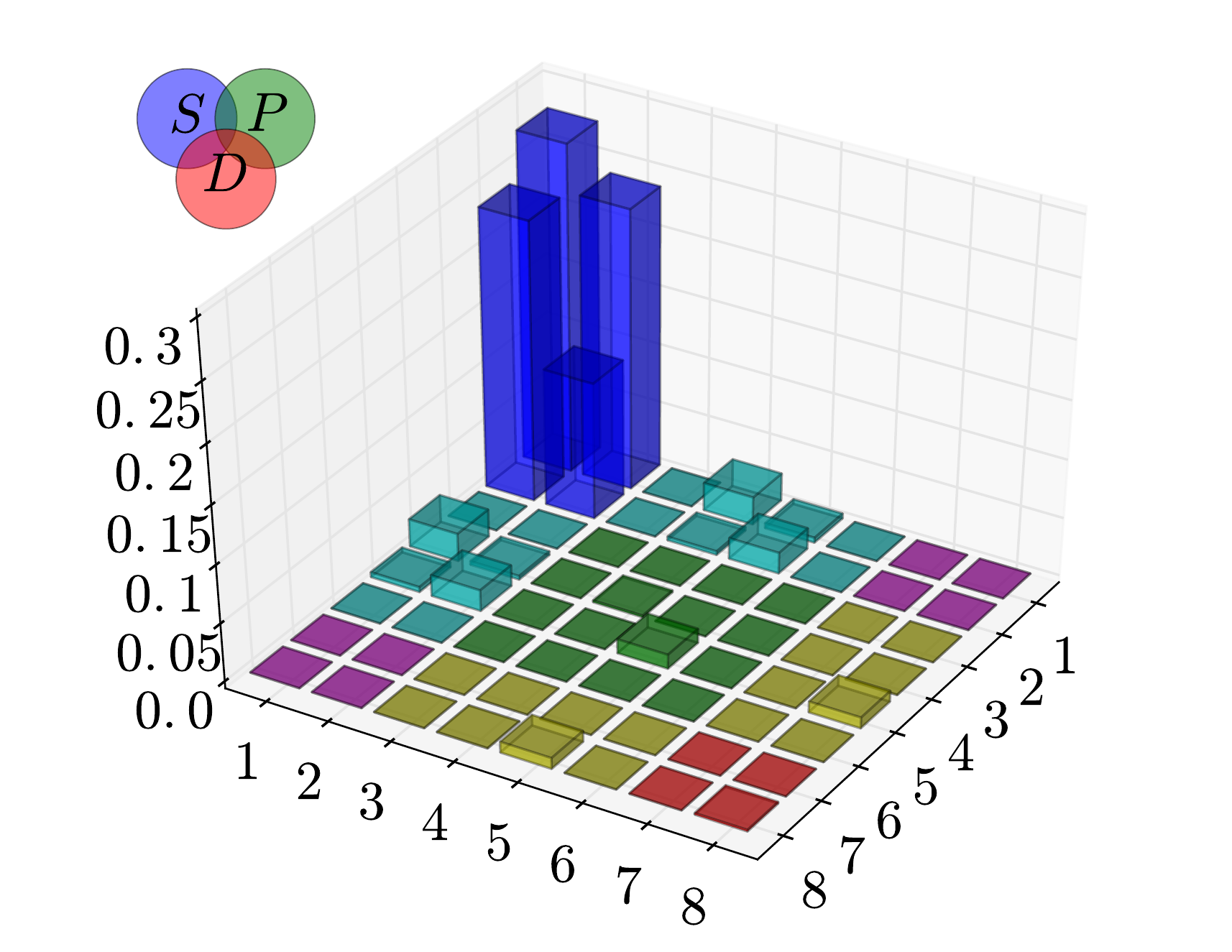}
  \includegraphics[width=.32\textwidth]{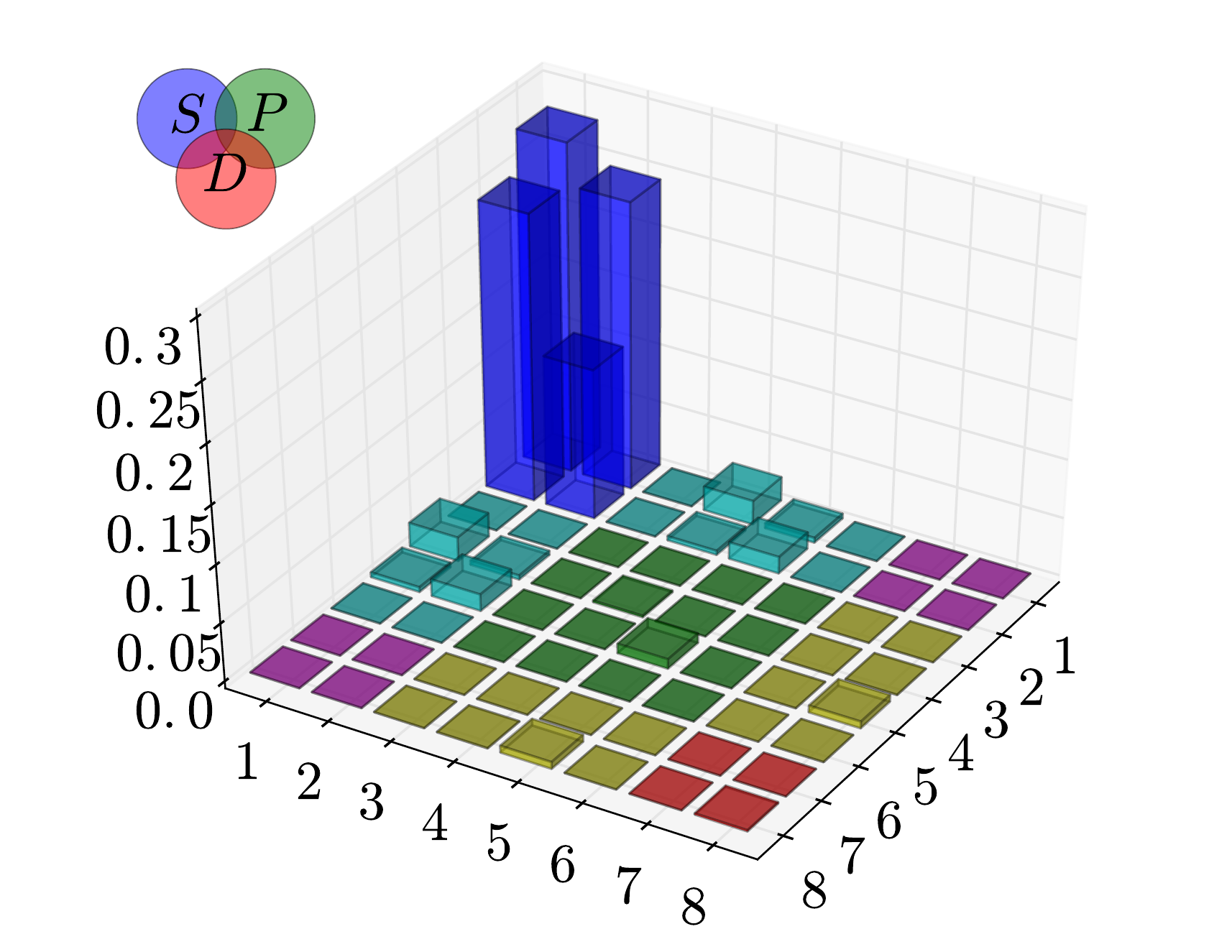}
  \includegraphics[width=.32\textwidth]{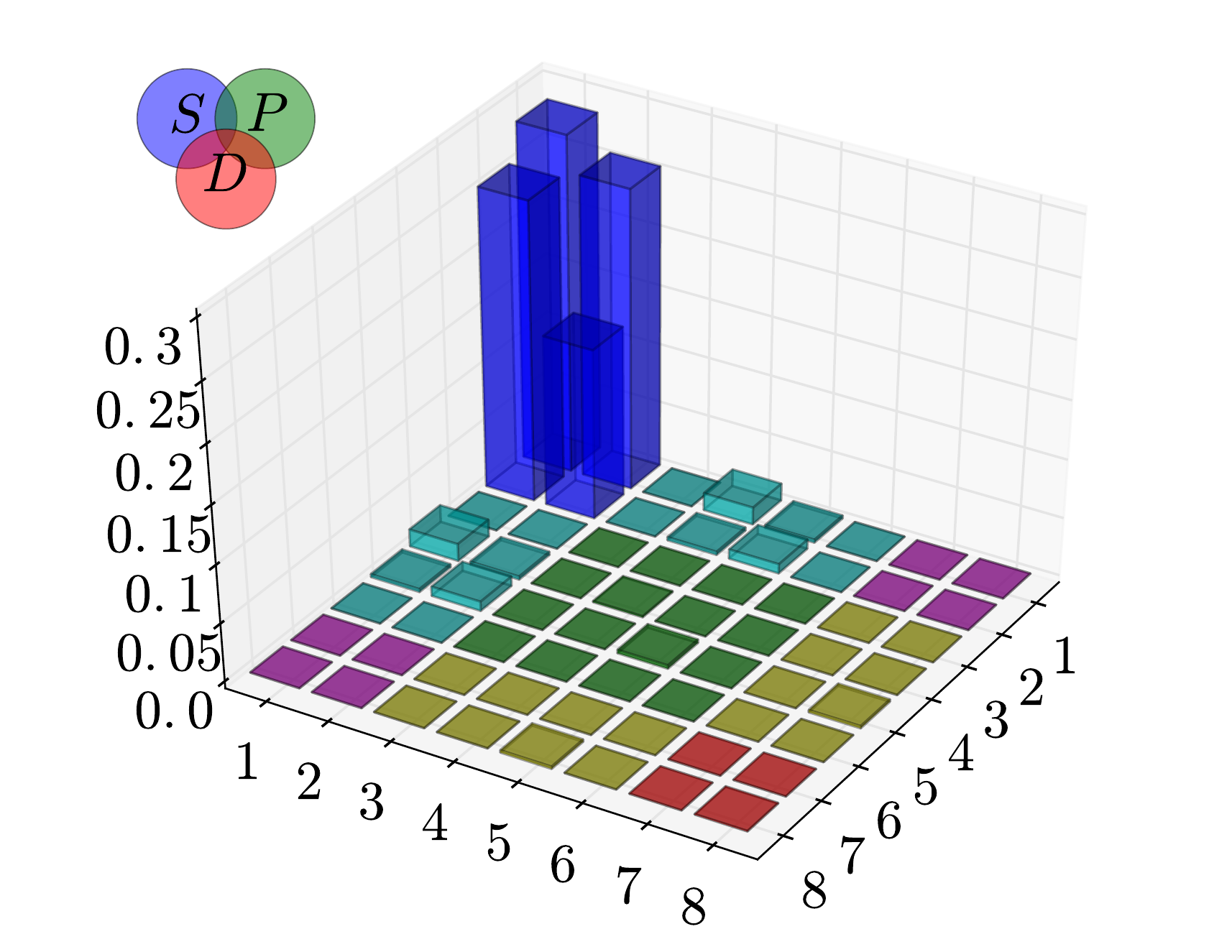}
\caption{\label{fig:orbital}
Orbital angular momentum contributions to the BSA's canonical norm for ground states. The rows in the figure correspond to 
key points on the graphs in Fig.~\ref{fig:mf} as follows (states in notation flavor-$n(J^{\mathcal{P}(\mathcal{C})})$): 
\emph{Upper row:} $\bar{u}u$-$0(0^{-+})$, $\bar{u}s$-$0(0^{-})$, $\bar{s}s$-$0(0^{-+})$;
\emph{Center row:} $\bar{u}u$-$0(0^{++})$, $\bar{u}s$-$0(0^{+})$, $\bar{s}s$-$0(0^{++})$;
\emph{Lower row:} $\bar{u}u$-$0(1^{--})$, $\bar{u}s$-$0(1^{-})$, $\bar{s}s$-$0(1^{--})$.}
\end{figure*}

However, there is a prominent case where $f$ provides a clear signal to distinguish conventional and quasi-exotic states:
excited charged pions. While herein we compute the pion's properties on the basis of equal $u$ and $d$ quark masses
as well as isospin symmetry, and also neglect electromagnetic effects on the different charge states, we can still conclude
that a charged excited pion connected to a conventional state should have $f\approx 1$ MeV as well, while quasi-exotic charged
pions have $f\approx 0.1$ MeV, i.\,e., one order of magnitude smaller. This can be seen from the first data point 
from the left in the lower left subfigure of Fig.~\ref{fig:mf}, which corresponds to a possible current-quark mass
combination for a realistic charged pion. Finding a state with such a quasi-exotic signature would thus immediately
signal the existence of exotic $0^{--}$ pseudoscalars of about the same mass. More precisely, the existence
of a quasi-exotic state implies the existence of two more associated exotic states.

The argument obviously works both ways, and so the existence of quasi-exotic states can also be inferred from the exotic case.
This is an interesting statement for the case of the $1^{-+}$ as shown in Fig.~\ref{fig:mf}.
The ground-state values of $f_\rho$ and $f_\phi$ are connected by a slow and steady increase, while the 
$f$ values for the conventional excited and quasi-exotic states differ by at least an order of magnitude
even in the strange case. 
We note that the zeros visible in the curve for the quasi-exotic state come from a sign change of $f$.

To close this section we would like to remark that the masses of the meson excitations discussed here are not well-anchored to exact results in QCD.
Thus, they can depend rather strongly on the model parameters and do, in general, not compare well to experimental data. 
Fine-tuning the parameters can be used to achieve a reasonable description of excited-state masses, but since this is not the focus of our argument herein, we refer the interested reader to our previous studies in this direction \cite{Hilger:2015ora} and references therein.

Another remark is that changing the model parameters might also change the behavior and role of states with regard to their level ordering, both in terms of quasi-exotic or conventional characteristic as well as the behavior of having real solutions of the homogeneous BSE or not.
However, as mentioned above, while states without real solutions of the homogeneous BSE have to be discarded in our present context, our picture and argument are still valid using the remaining states.

\section{Orbital angular momentum\label{sec:orbital}}

While $l$ is not a Lorentz invariant, the covariants in Eq.~(\ref{eq:tbase}) can be identified
with orbital angular momentum $l$ associated with the relative $\bar{q}q$ momentum
in the meson's rest frame \cite{Bhagwat:2006xi}. 
Terms unexpected in a quark-model setup appear due to the fully covariant amplitude, such as $P$-wave 
components in the pseudoscalar or vector BSA or $S$-wave components in the scalar one. 

\begin{figure*}[t]
  \includegraphics[width=.32\textwidth]{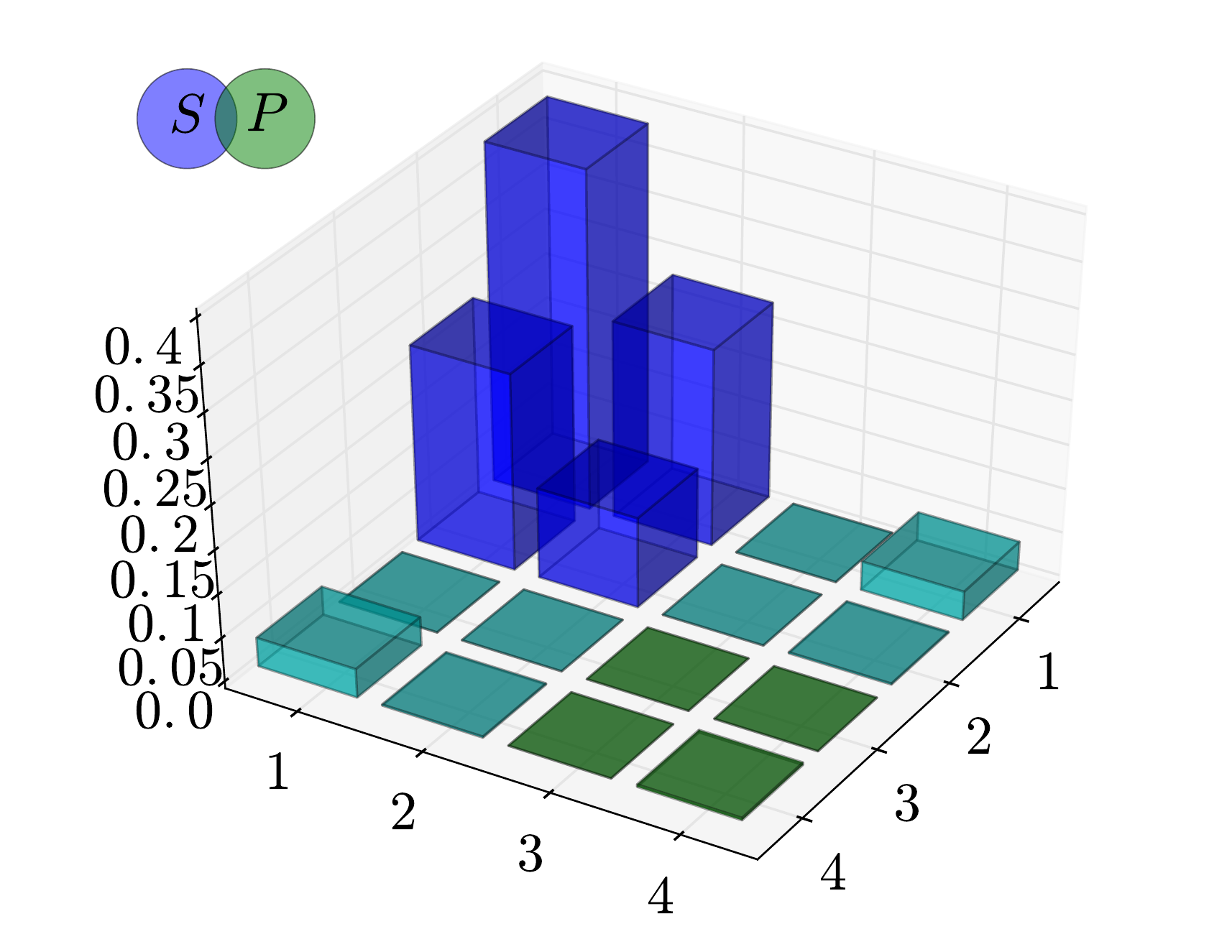}
  \includegraphics[width=.32\textwidth]{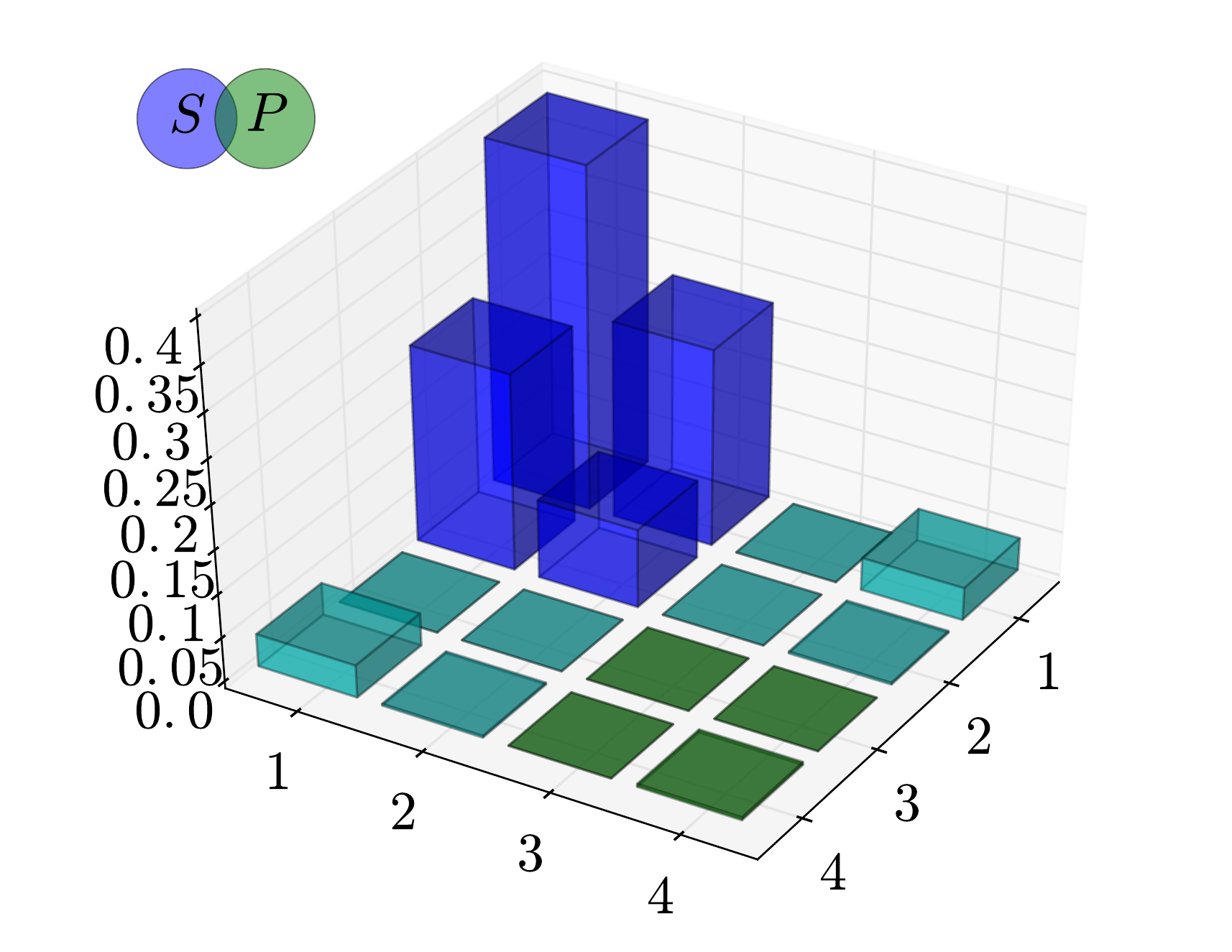}
  \includegraphics[width=.32\textwidth]{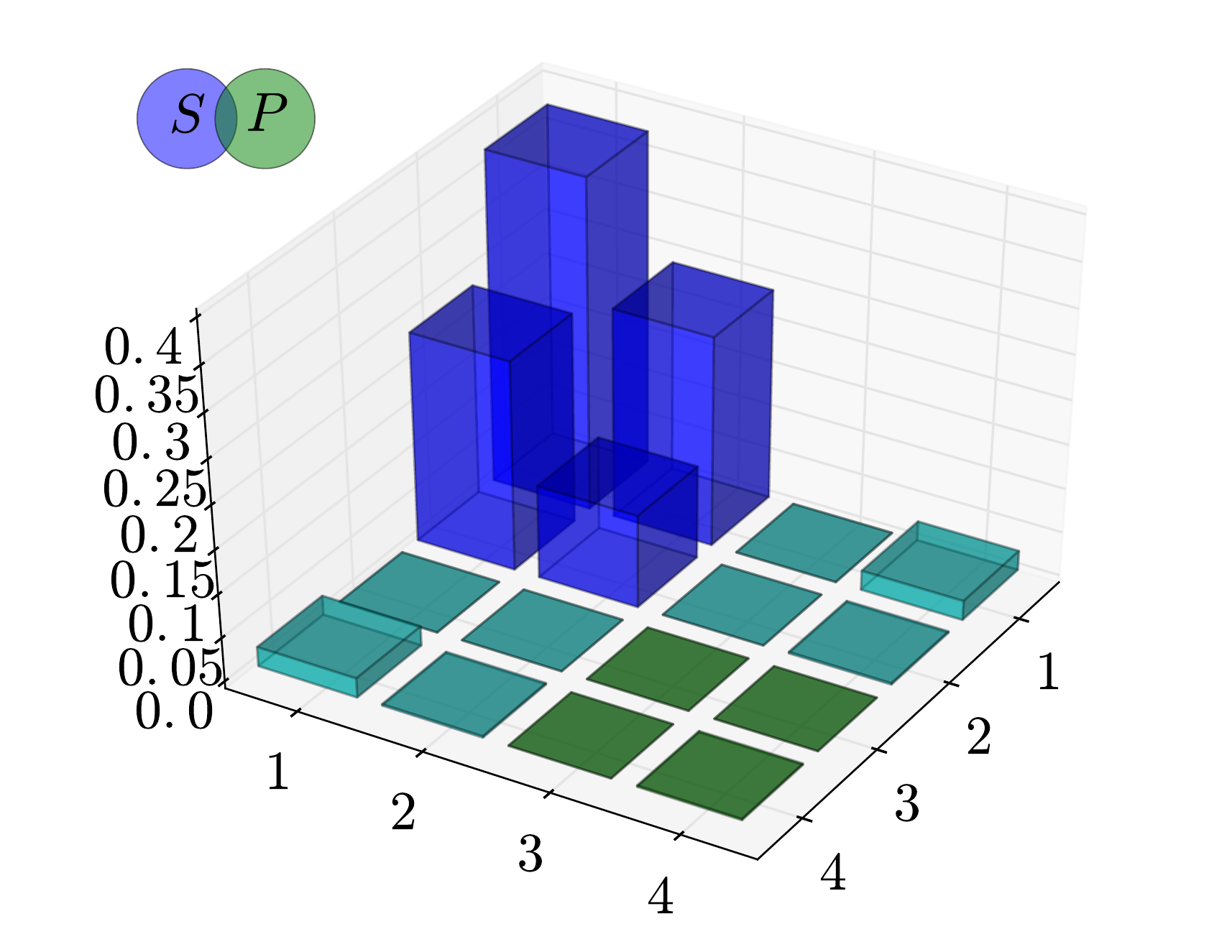}
  \includegraphics[width=.32\textwidth]{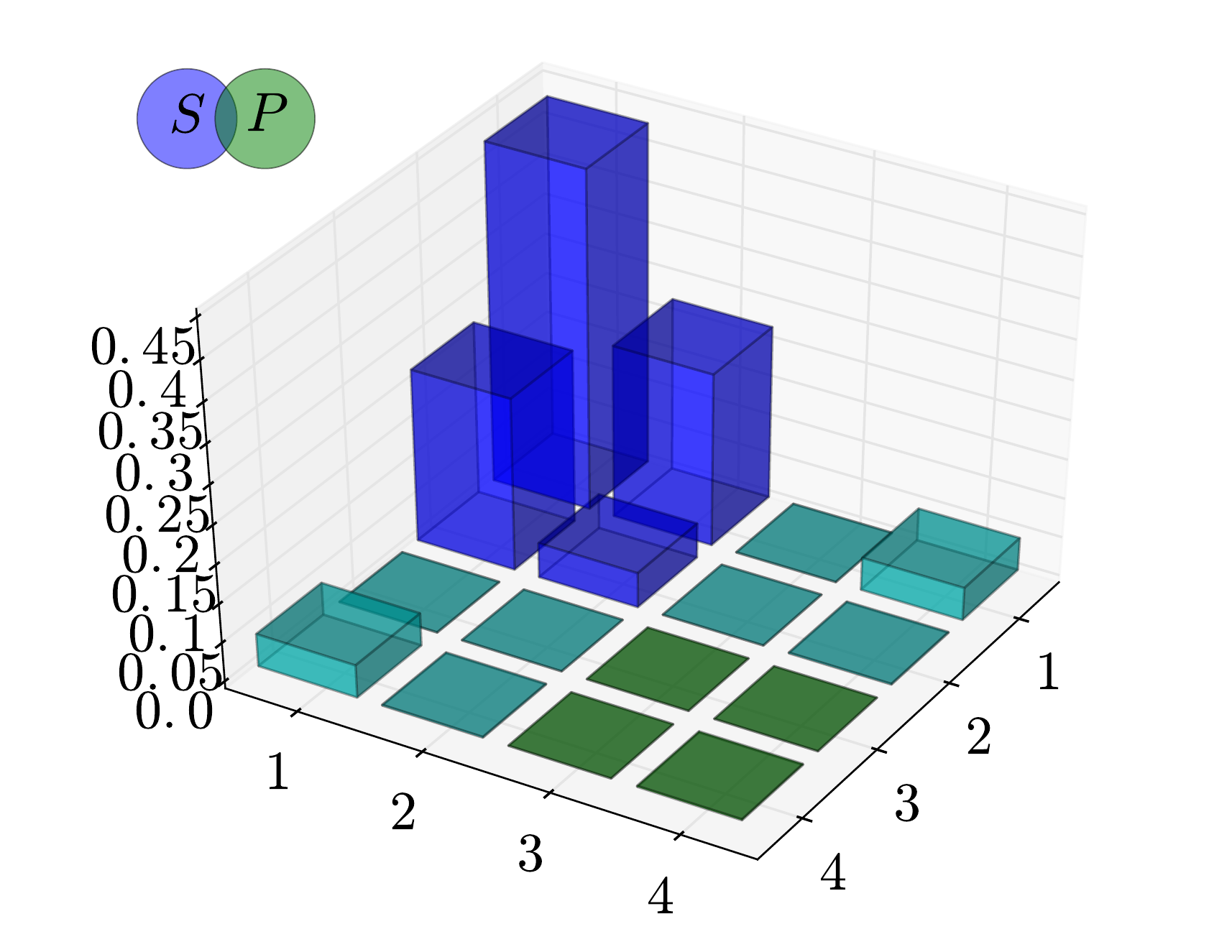}
  \includegraphics[width=.32\textwidth]{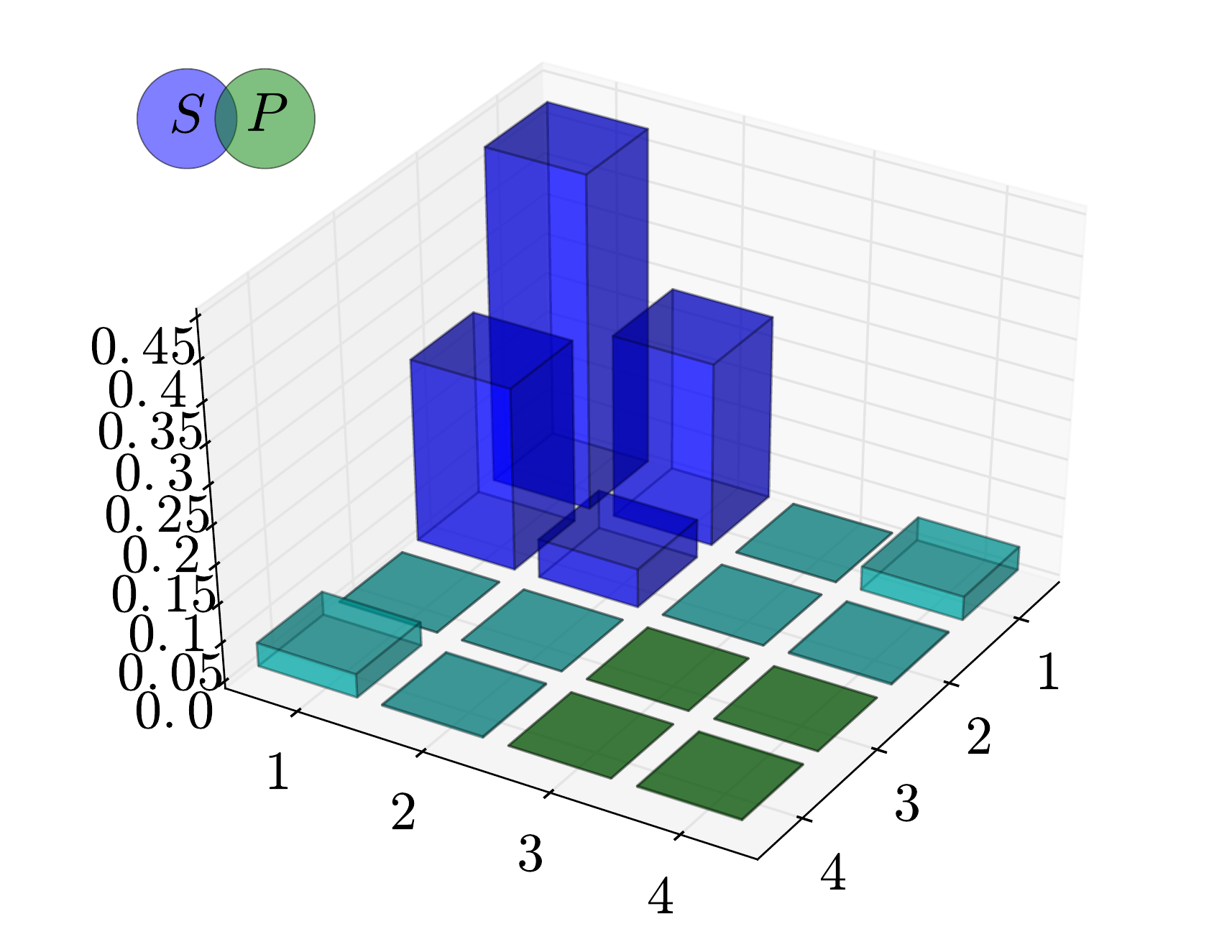}
  \includegraphics[width=.32\textwidth]{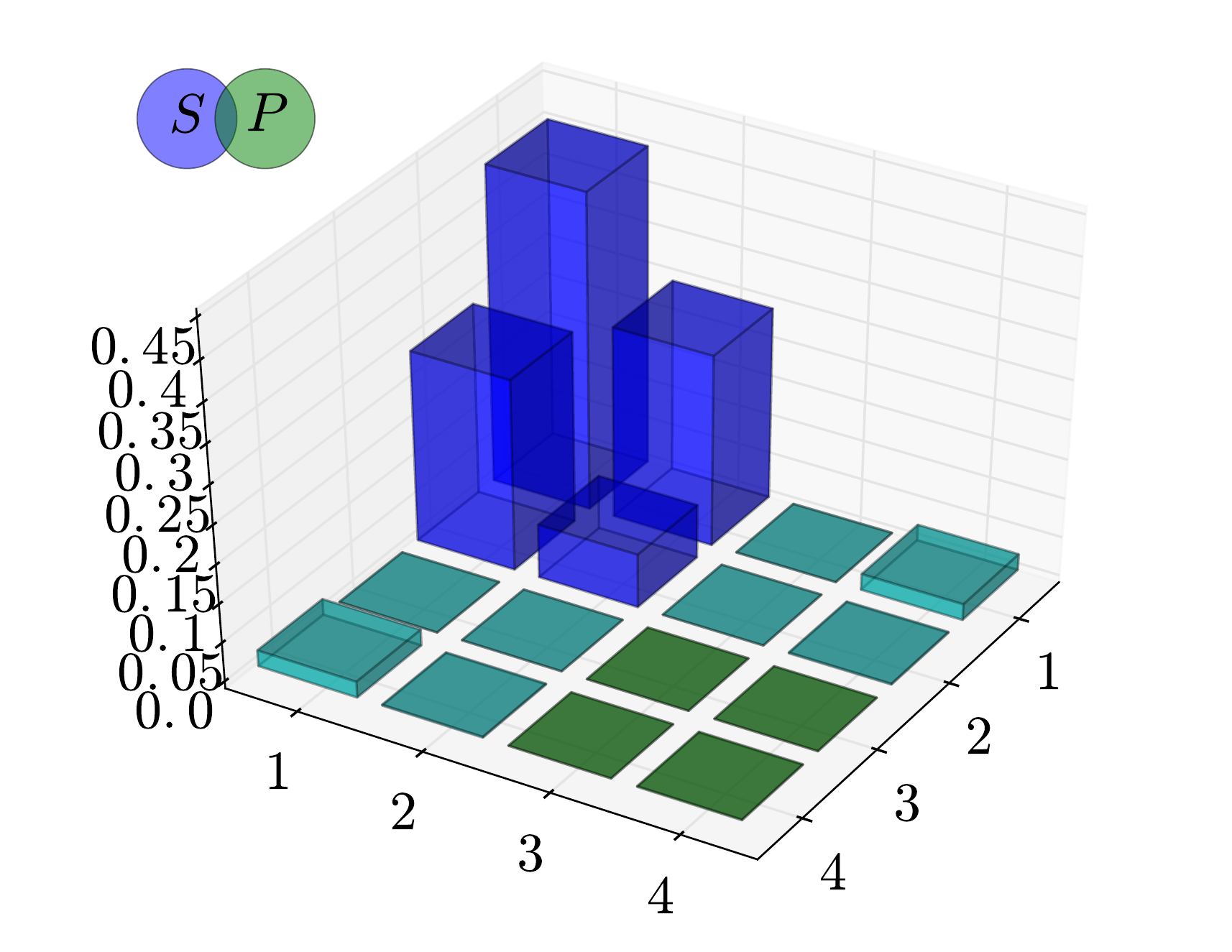}
  \includegraphics[width=.32\textwidth]{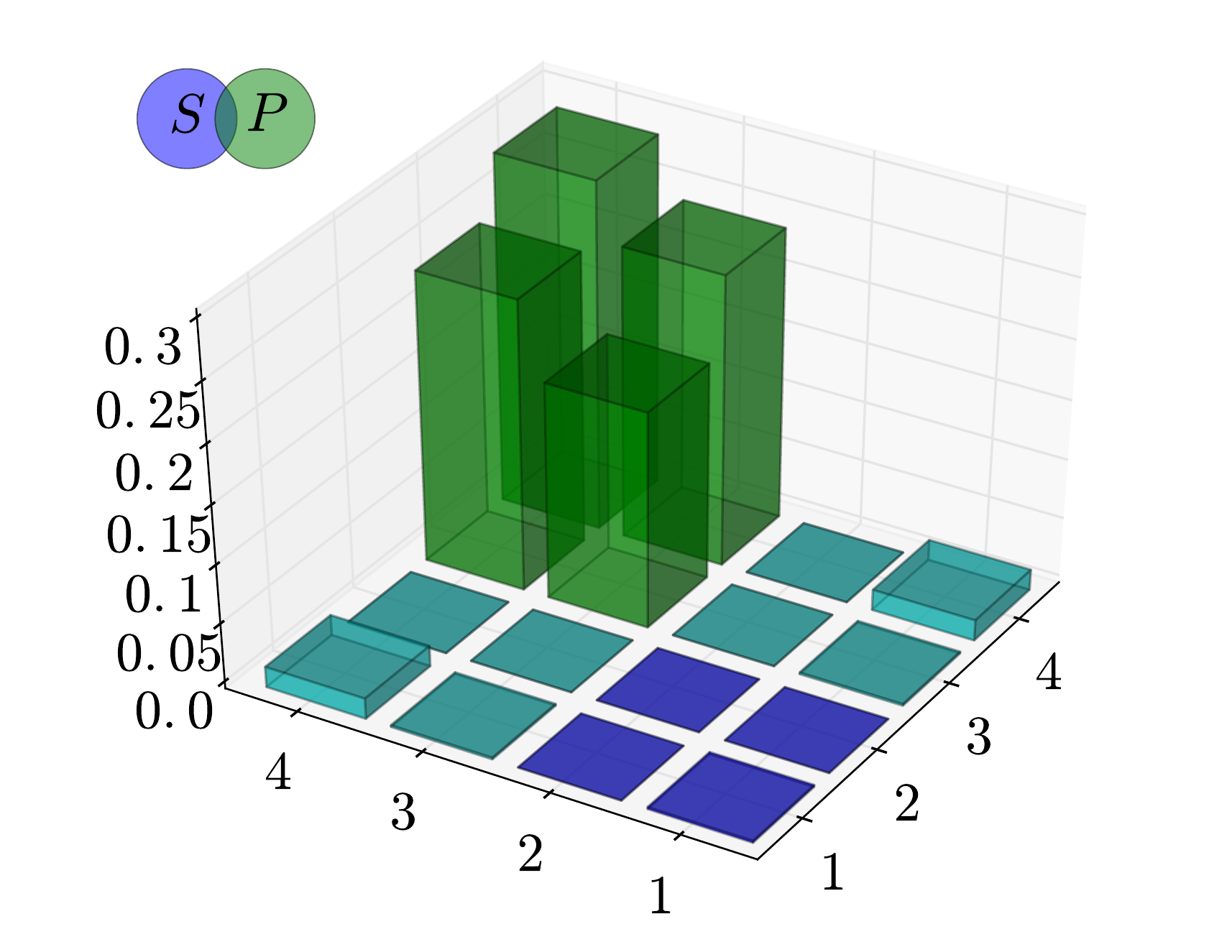}
  \includegraphics[width=.32\textwidth]{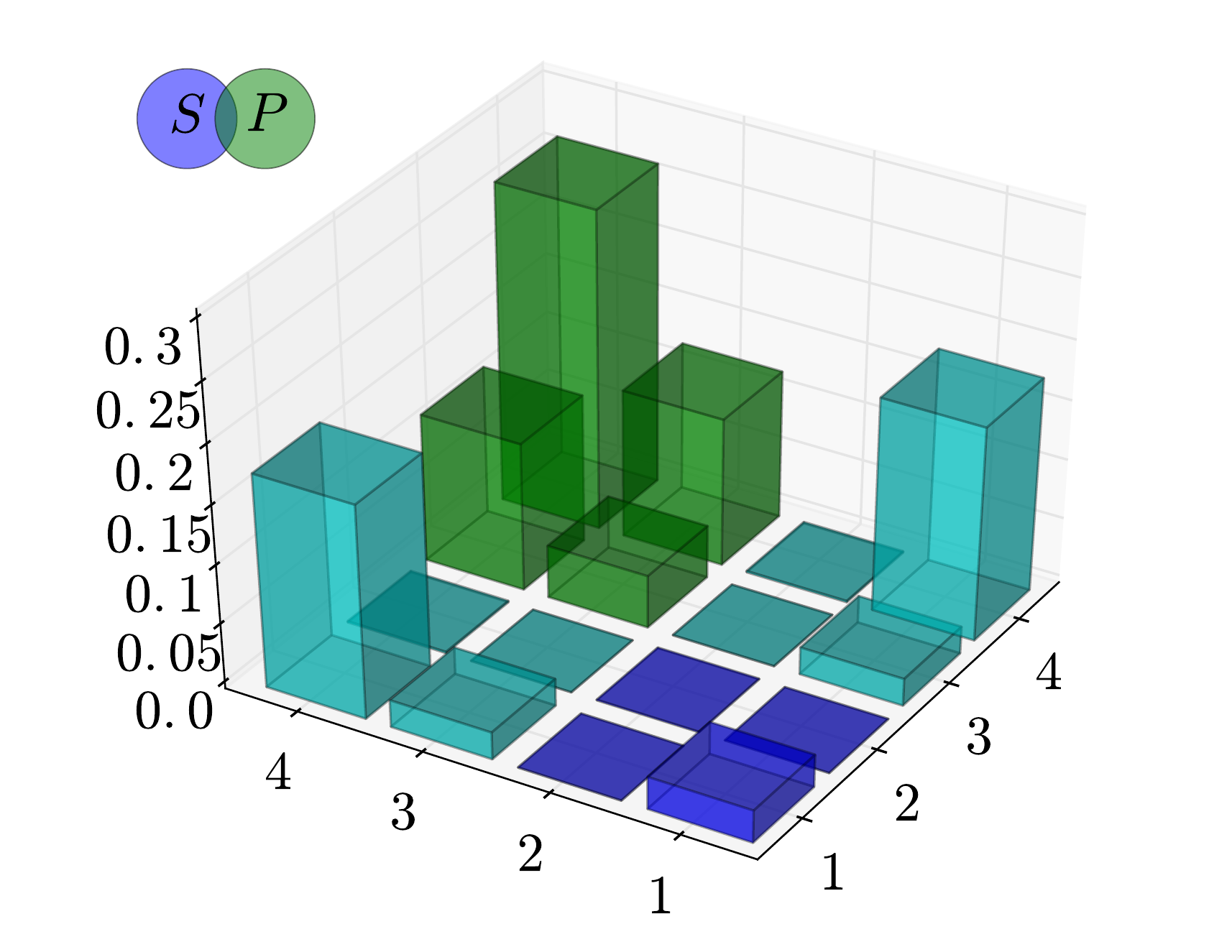}
  \includegraphics[width=.32\textwidth]{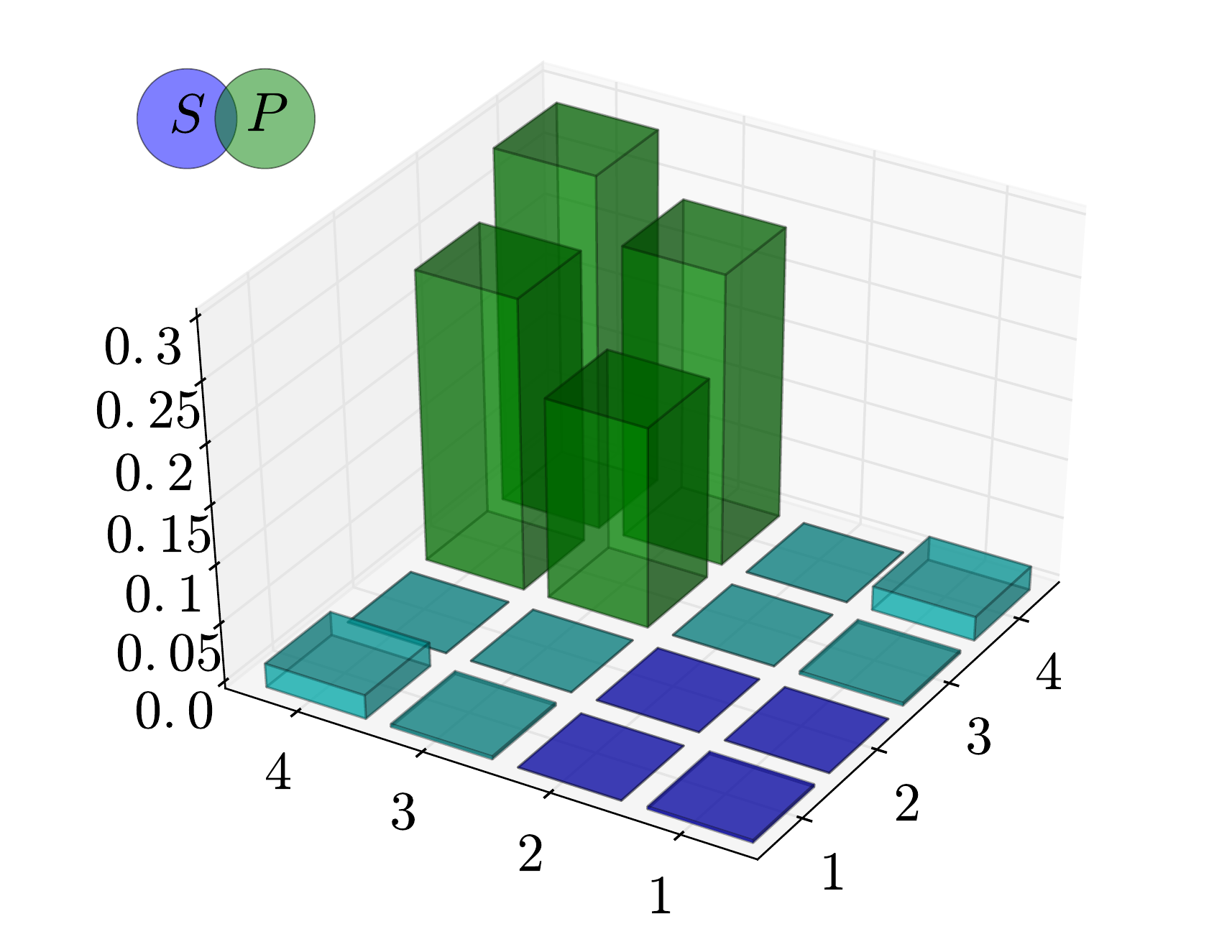}
  \includegraphics[width=.32\textwidth]{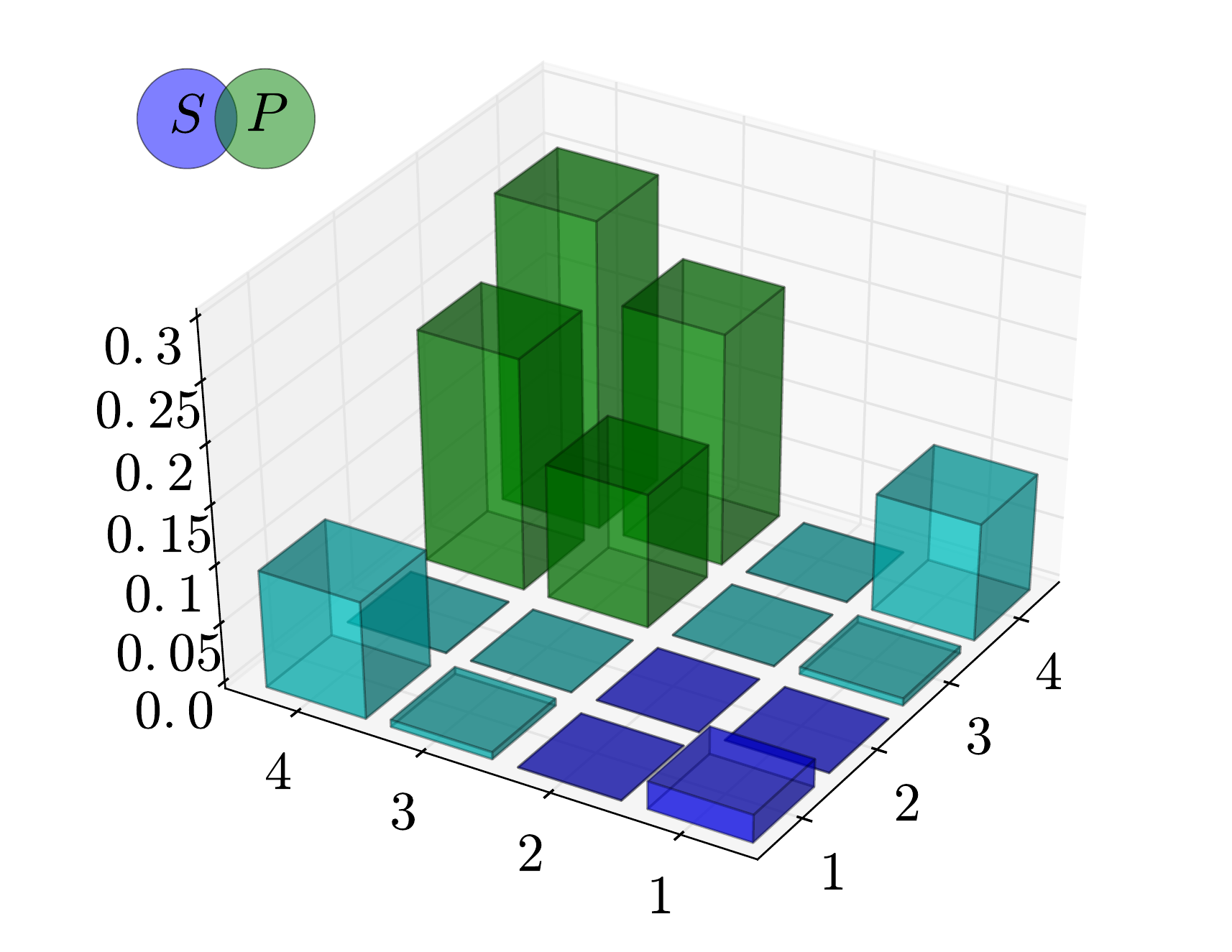}
  \includegraphics[width=.32\textwidth]{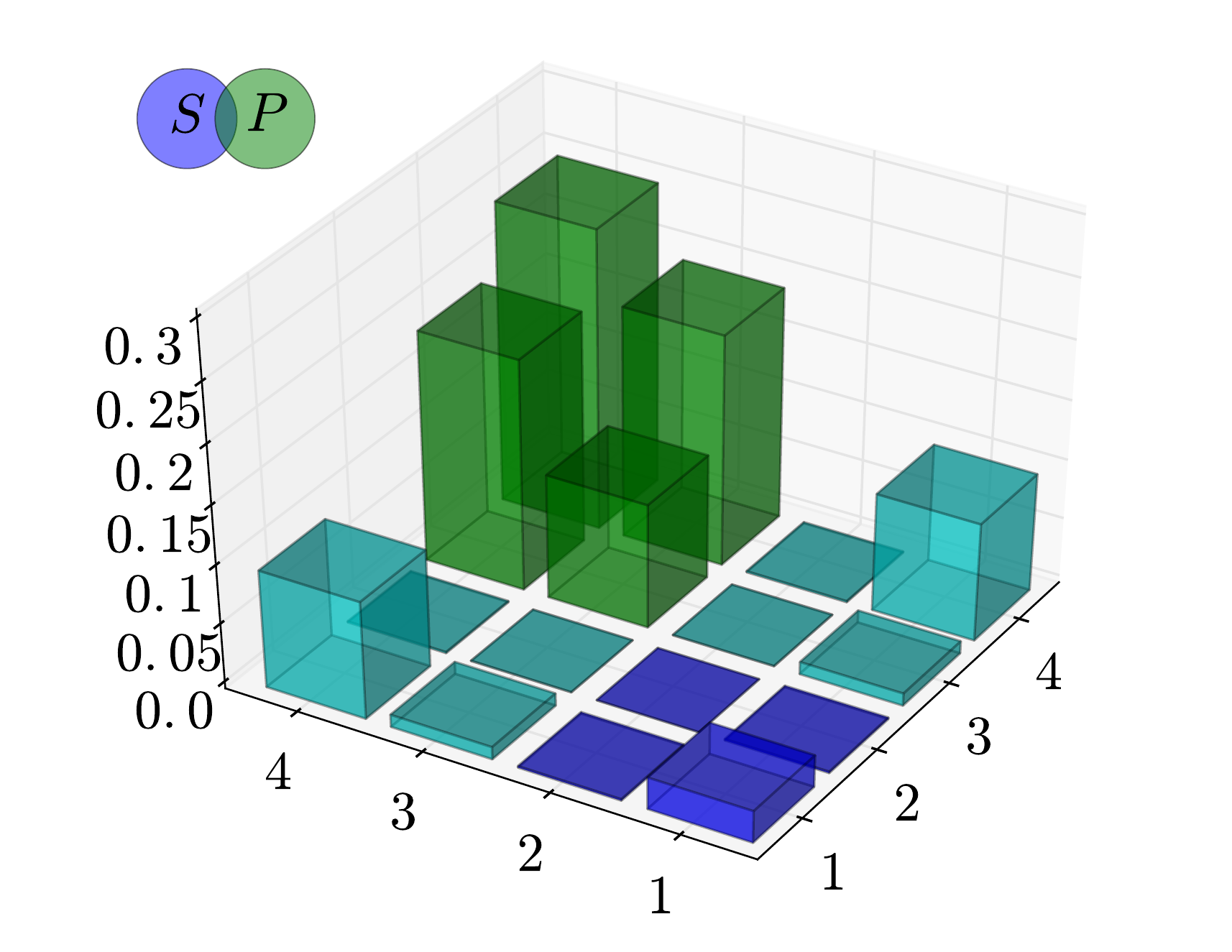}
  \includegraphics[width=.32\textwidth]{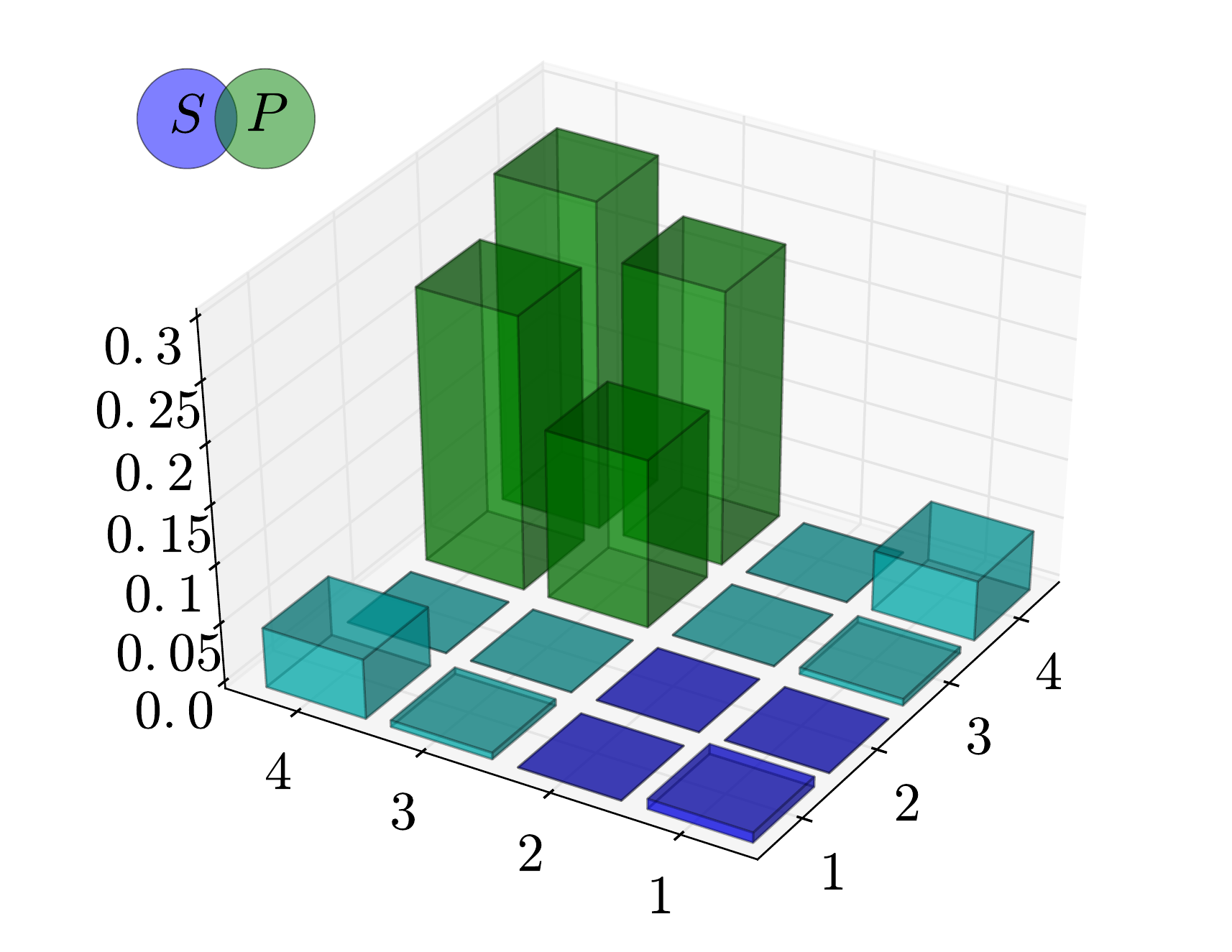}
\caption{\label{fig:orbital-e}
Orbital angular momentum contributions to the BSA's canonical norm for conventional and exotic excited quarkonia with $J=0$. 
The rows in the figure correspond to 
key points on the graphs in Fig.~\ref{fig:mf} as follows (states in notation flavor-$n(J^{\mathcal{P}(\mathcal{C})})$): 
\emph{Upper row:} $\bar{u}u$-$1(0^{-+})$, $\bar{u}s$-$2(0^{-})$, $\bar{s}s$-$1(0^{-+})$;
\emph{Upper center row:} $\bar{u}u$-$0(0^{--})$, $\bar{u}s$-$1(0^{-})$, $\bar{s}s$-$0(0^{--})$;
\emph{Lower center row:} $\bar{u}u$-$1(0^{++})$, $\bar{u}s$-$2(0^{+})$, $\bar{s}s$-$1(0^{++})$;
\emph{Lower row:} $\bar{u}u$-$0(0^{+-})$, $\bar{u}s$-$1(0^{+})$, $\bar{s}s$-$0(0^{+-})$.}
\end{figure*}

Here, we focus on the comparison of ground states and their excitations including the open-flavor case as
they are contained in Fig.~\ref{fig:mf}. The $l$-content of several key states
is presented in Figs.~\ref{fig:orbital} and \ref{fig:orbital-e} as well as in Tab.~\ref{tab:orbital} in the following way: the covariants in Eq.~(\ref{eq:tbase}) are
numbered $1$ to $4$: $1$ and $2$ correspond to $S$-wave, $3$ and $4$ are $P$-wave. For the vector
meson each of these covariants are combined with the two four-vectors $\gamma^\mu$ and $k^\mu$, arriving
at a set of eight vector covariants (for details, see, e.\,g., Ref.~\cite{Krassnigg:2010mh}), thus numbered $1$ to $8$.
In the vector case, covariants $1$ and $2$ correspond to $S$-wave, $3$ - $6$ are $P$-wave, and $7$ and $8$ are $D$-wave.

After solving the homogeneous BSE, we canonically normalize the BSA and explicitly extract the contributions
from each combination of covariants to the norm. The squares of these contributions are plotted in 
Figs.~\ref{fig:orbital} and \ref{fig:orbital-e} with their sum normalized to one. Full details on this 
kind of construction, the complete sets of covariants as well as the basis for the interpretation in 
terms of $l$ can be found in Ref.~\cite{Hilger:2015ora}.

We start with the discussion of ground states presented in Fig.~\ref{fig:orbital}:  Next to each other in each row of the figure,
we present the key states along the ground-state trajectories as outlined in Fig.~\ref{fig:quasiexotic} and plotted
in Fig.~\ref{fig:mf} for $J^{\mathcal{P}\mathcal{C}}=0^{-+}$, $0^{++}$, and $1^{--}$. In this way, the three
columns in Fig.~\ref{fig:orbital} contain the flavor combinations $\bar{u}u$, $\bar{u}s$, and $\bar{s}s$, respectively.
For easy reference, all numbers corresponding to the bar heights in the figures are collected in Tab.~\ref{tab:orbital}.

It is obvious immediately that changes along the trajectory $\bar{u}u \rightarrow \bar{u}s \rightarrow \bar{s}s$ are 
very small. For example, the pion is more than $80$\% $S$-wave and negligible pure $P$-wave.
The same picture holds for the $K$ and a theoretical pure flavor $\bar{s}s$ pseudoscalar ground state. It is important to
note here that we always deal with ideally mixed flavor states due to the absence of flavor-mixing or -changing contributions in
the RL BSE interaction kernel. This does not mean, however, that an investigation of flavor-mixed states would be impossible
at our level of sophistication, since one can always perform mixing at the hadronic level \cite{Holl:2004un}. In fact,
we can indeed argue on the basis of our results that any particular mixture of $\bar{u}u$ and $\bar{s}s$ (and, of course, 
away from the isospin-symmetric limit also $\bar{d}d$) should have an orbital angular momentum content very similar 
to any of the pure $\bar{f}f$ states for any of the flavors $f=u,d,s$. 

Similarly, in the scalar and vector cases, the ground states do not only appear in the $l$-configuration
expected from the quark model \cite{Hilger:2015ora}, but also stay almost unchanged along their trajectories.
A few key examples are the $a_0(980)$ with only $7$\% $S$-wave, $45$\% $P$-wave, and $48$\% 
mixed contributions, representative of also the $\kappa$, as well as the $\rho$ with $87$\% $S$-wave, negligible pure 
$P$- and $D$-wave parts, and a total of $12$\% of various mixed contributions, also representative of the $K^*$ and the $\phi$.

The excited-state results are shown in Fig.~\ref{fig:orbital-e} for both conventional and exotic excited quarkonia with
$J^{\mathcal{P}}=0^{-}$ and $0^{+}$. In particular, the states are extracted in accordance to those shown in Fig.~\ref{fig:mf}
such that the first and third rows in Fig.~\ref{fig:orbital-e} contain conventional $J^{\mathcal{P}\mathcal{C}}$ excitations
while the second and forth rows show the transitions from exotic to quasi-exotic back to exotic states along the 
$\bar{u}u \rightarrow \bar{u}s \rightarrow \bar{s}s$ trajectory. Note that on such a trajectory the excitation quantum number
$n$ changes due to the increased number of states in a $J^{\mathcal{P}}$ channel, which contains both signs of 
$\mathcal{C}$ in the respective $J^{\mathcal{P}\mathcal{C}}$ channels.

\begin{table}[h]
\caption{Orbital angular momentum content of states shown in Figs.~\ref{fig:orbital} and \ref{fig:orbital-e}. 
Numbers are given in \%. Question marks parentheses indicate that the mass of such a state, which would likely be included
in the name, is not experimentally known yet as is the case also for other states listed here, e.\,g., the $b_0$, which is
an isovector exotic scalar. 
A $\prime$ indicates an excitation analog to a radial excitation in
the quark model. Flavor subscripts denote ideally mixed components.\label{tab:orbital}}
\begin{tabular*}{\columnwidth}{l@{\extracolsep{\fill}}rrrrrrr}
\hline\noalign{\smallskip}
Name & $n(J^{PC})$       & $S$   & $S$-$P$ & $P$ & $P$-$D$ & $D$ & $S$-$D$ \\ \noalign{\smallskip}\hline\noalign{\smallskip}
$\pi$ & $0(0^{-+})$      & $81.8$ & $18.2$ & $0.0$ & $-$ & $-$ & $-$ \\ 
$K$ & $0(0^{-})$         & $85.0$ & $14.9$ & $0.0$ & $-$ & $-$ & $-$ \\ 
$\eta_{\bar{s}s}$ & $0(0^{-+})$        & $90.8$ & $9.1$ & $0.0$ & $-$ & $-$ & $-$ \\ 
$a_0(980)$ & $0(0^{++})$ & $6.6$ & $48.4$ & $45.0$ & $-$ & $-$ & $-$ \\
$\kappa$ & $0(0^{+})$    & $5.5$ & $43.0$ & $51.4$ & $-$ & $-$ & $-$ \\ 
$f_{0\bar{s}s}$ & $0(0^{++})$        & $2.7$ & $28.5$ & $68.8$ & $-$ & $-$ & $-$ \\
$\rho$ & $0(1^{--})$     & $86.7$ & $9.8$ & $1.4$ & $1.9$ & $0.1$ & $0.0$ \\
$K^*$ & $0(1^{-})$     & $89.1$ & $8.6$ & $1.0$ & $1.2$ & $0.1$ & $0.0$ \\
$\phi$ & $0(1^{--})$     & $93.5$ & $5.6$ & $0.4$ & $0.4$ & $0.0$ & $0.0$ \\
$\pi(1300)$   & $1(0^{-+})$   & $92.9$ & $6.8$ & $0.3$ & $-$ & $-$ & $-$ \\
$K(1460)$ & $2(0^{-})$   & $91.9$ & $7.8$  & $0.3$ & $-$ & $-$ & $-$ \\ 
$\eta\prime_{\bar{s}s}$ & $1(0^{-+})$        & $95.1$ & $4.7$ & $0.2$ & $-$ & $-$ & $-$ \\ 
$\rho_0$      & $0(0^{--})$   & $91.3$ & $8.6$ & $0.1$ & $-$ & $-$ & $-$ \\
$K(?)$ & $1(0^{-})$         & $93.6$ & $6.3$  & $0.1$ & $-$ & $-$ & $-$ \\ 
$\omega_{0\bar{s}s}$      & $0(0^{--})$   & $95.6$ & $4.3$ & $0.1$ & $-$ & $-$ & $-$ \\
$a_0(1450)$ & $1(0^{++})$ & $0.2$ & $3.8$ & $96.1$ & $-$ & $-$ & $-$ \\
$K_0^*(1430)$ & $2(0^{+})$& $2.8$ & $40.9$ & $56.3$ & $-$ & $-$ & $-$ \\ 
$f\prime_{0\bar{s}s}$ & $1(0^{++})$        & $0.2$ & $4.6$ & $95.2$ & $-$ & $-$ & $-$ \\
$b_0$      & $0(0^{+-})$   & $2.4$ & $21.2$ & $76.3$ & $-$ & $-$ & $-$ \\
$K_0^*(?)$ & $1(0^{+})$         & $3.0$ & $22.3$ & $74.7$ & $-$ & $-$ & $-$ \\ 
$h_0$      & $0(0^{+-})$   & $0.9$ & $11.3$ & $87.8$ & $-$ & $-$ & $-$ \\
\noalign{\smallskip}\hline\noalign{\smallskip}
\end{tabular*}
\end{table}

It is remarkable that there is no strong difference between the conventional and exotic excitations in both 
cases, i.\,e., the exotic character of the corresponding excited states in our approach does not come
about via some kind of excitation in orbital-angular 
momentum corresponding to the $\bar{q}q$ relative momentum. Rather, it seems to be a mechanism akin to 
a radial excitation in the quark model. We note again at this point that gluonic excitation mechanisms
are not built in explicitly, but implicitly in our approach.

\begin{figure*}[t]
  \includegraphics[width=\textwidth]{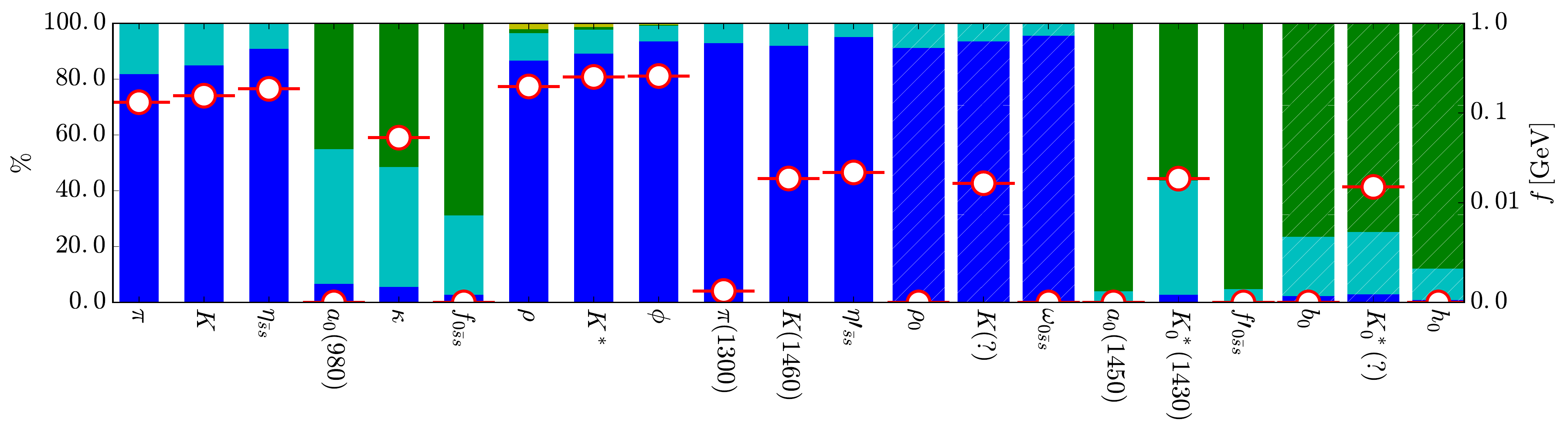}
\caption{\label{fig:oov}
Orbital angular momentum contributions. Columns labeled by experimental states, where available.
Colors: $S$ (blue), $P$ (green), $D$ (red), $S$-$P$-mix (cyan), $S$-$D$-mix (magenta), $P$-$D$-mix (yellow).
(Quasi-)exotic bands are wider and hatched.
\emph{Overlay} (right vertical axis, linear below $0.01$ GeV): leptonic decay constants (red filled circles with lines). }
\end{figure*}

In Fig.~\ref{fig:oov} we plot and overview of the $l$ contributions together with the values for $f$ on a logarithmic scale
for all states shown in Figs.~\ref{fig:orbital} and \ref{fig:orbital-e}.

\section{Charmed strange sector}

Open-charm mesons are of particular DCSB relevance \cite{Hilger:2008jg,Hilger:2011cq,Buchheim:2014rpa,Rapp:2011zz,Buchheim:2015xka}.
We add a small set of results containing the $D_s$ meson and its first excitation using the slightly
different model interaction of Ref.~\cite{Alkofer:2002bp}. 
In particular, this interaction does not have the correct asymptotic behavior of the QCD running coupling, which makes a re-determination of quark masses necessary, which we achieve analogously to the fitting scheme described above for the light-strange meson results. 
With $\omega=0.6$ GeV,  $D=1$ GeV${^2}$, the current-quark masses $m_u = 0.0045$ GeV, $m_s = 0.11$ GeV, and $m_c = 0.97$ GeV are 
fitted to the $\pi$, $K$, and $D$-meson masses to yield $M_\pi=0.137$, $f_\pi=0.129$, $M_K=0.499$, and $f_K=0.157$ GeV 
as well as $M_D = 1.868$ and $f_D = 0.268$ GeV compared
to the experimental $M_{D^+} = 1.86961(9)$ and $f_{D^+} = 0.2046(50)$ GeV \cite{Olive:2014rpp}.

\begin{table}[h] 
\centering
\caption{\label{tab:cs}Charmed strange results in GeV.}
\begin{tabular}{ccccccc}
 & \multicolumn{2}{c}{$\bar{s}s$} & \multicolumn{2}{c}{$\bar{c}s$} & \multicolumn{2}{c}{$\bar{c}c$}\\ \cline{2-3}\cline{4-5}\cline{6-7}
$J^{\mathcal{P}(\mathcal{C})}$ & $M$     & $|f|$ 	& $M$     & $|f|$   &$M$      & $|f|$ \\ \hline
$0^{-(+)}$ & $0.698$ & $0.187$ & $1.888$ & $0.268$ & $2.710$ & $0.342$ \\ 
$0^{-(-)}$ & $1.427$ &	$0.000$	& $2.357$ & $0.180$ & $3.113$ & $0.000$ 
\end{tabular}
\end{table}

The results for combinations of $c$ and $s$ flavors are given in Tab.~\ref{tab:cs}.
Experimentally one has $M_{D_s^+} = 1.9683(1)$ and $f_{D_s^+} = 0.2575(46)$ GeV.
The first radial excitation is connected to exotic pseudoscalars for $\bar{c}c$ and $\bar{s}s$, which shows
that for large asymmetries regarding the quark masses inside the meson, $f$ is not necessarily a reliable means to 
distinguish whether experimental states are quasi-exotic or not.

\section{Conclusions and outlook\label{sec:conclusions}}

In summary, we argue for the existence of open-flavor analogons of exotic mesons from a quark-bilinear covariant BSA on the 
basis of the following points: 
\begin{enumerate}
\item there is no conceptual difference between the construction of exotic and non-exotic 
quark-bilinear meson states within a Poincare covariant bound-state approach
\item there is no conceptual difference in the 
construction of an equal-flavor and an open-flavor state
\item based on our OAMD results, the inherent character of such 
states is not altered along quark-mass changing trajectories depicted in Fig.~\ref{fig:quasiexotic}
\item it is reasonable to assume continuity of meson spectra with respect to variations of the quark mass
\end{enumerate}
It is important to note here that including hadronic decay channels in a more sophisticated truncation would certainly have qualitative impact on the results.
However, the central results of our study should be robust with regard to such an extension, since they mainly regard how states with certain properties appear together or in relation to each other. 
In addition, since our anchoring limits and relations provide exact results in QCD and this should be also the case in any more sophisticated truncation of the DSBSE approach including those where hadronic channels are taken into account, the robustness of the anchor should translate well onto the key results of any more sophisticated study.

To the best of our knowledge, we describe and analyse such quasi-exotic states, as we term them due to their missing
restriction in the decisive quantum number $\mathcal{C}$, for the first time:
by continuity with respect to the quark masses they do not exist in the quark model, and
a setup like ours has not yet been explored before.

In particular, we have shown how and if, on the basis of a covariant meson amplitude, quasi-exotic mesons can be identified
by the order of magnitude of their leptonic decay constants. While the prime example is that of charged
quasi-exotic pion excitations, there can be other cases where such an identification is clear and would
provide evidence for quasi-exotic states in the open-flavor meson spectrum with clear connections to their
exotic quarkonium partners.

Orbital angular $\bar{q}q$ momentum, on the other hand, provides no clues to discern such states
in our study. However, the orbital-angular-momentum decomposition of the states investigated here
shows clear similarities along the $\bar{u}u \rightarrow \bar{u}s \rightarrow \bar{s}s$ trajectory. 
This supports the concept of flavor-mixing mechanisms executed at the hadronic level in general, but in
particular in the DSBSE approach. In addition it clearly demonstrates the conceptual similarity of 
exotic and quasi-exotic quark bilinear meson states.

As a result of the correspondence shown between exotic $J^{\mathcal{P}\mathcal{C}}$ and quasi-exotic $J^{\mathcal{P}}$ 
states, for each flavor combination there should be more states at and above the lowest (quasi)exotic 
meson mass than expected in the traditional quark model. 
Next steps to provide a better grasp on quasi-exotic states include calculations of their
hadronic decay properties along the lines of \cite{Jarecke:2002xd,Mader:2011zf}.

\begin{acknowledgements}
We acknowledge helpful conversations with M. Gomez-Rocha.
This work was supported by the Austrian Science Fund (FWF) under project no.\ P25121-N27.
\end{acknowledgements}



%


\end{document}